\documentclass[aps,pra,showkeys,twocolumn,superscriptaddress]{revtex4-2}
\usepackage{amsmath, amssymb}
\usepackage{mathrsfs}
\usepackage{array}
\usepackage{booktabs}
\usepackage{tabu}
\usepackage{dcolumn}
\usepackage{amsmath}
\usepackage{amsfonts}
\usepackage{float}
\usepackage{amssymb}
\usepackage{graphicx,color}
\usepackage{tikz}
\usepackage[colorlinks={true}]{hyperref}
\hypersetup{colorlinks=true,linkcolor=red,citecolor=blue,urlcolor=blue}
\usepackage{graphicx}
\usepackage{subfigure}
\usepackage{graphicx}% Include figure files
\usepackage{dcolumn}% Align table columns on decimal point
\usepackage{bm}% bold math
\usepackage{pstricks}
\usepackage{braket}
\usepackage{orcidlink}

\def\be{\begin{equation}}
  \def\ee{\end{equation}}
\def\bea{\begin{eqnarray}}
\def\eea{\end{eqnarray}}
\def\f{\frac}
\def\n{\nonumber}
\def\l{\label}
\def\p{\phi}
\def\o{\over}
\def\R{\hat{\rho}}
\def\pa{\partial}
\def\om{\omega}
\def\na{\nabla}
\def\P{\Phi}
\begin{document} 
\title{
Coherence, Transport, and Chaos in 1D Bose-Hubbard Model: \\Disorder vs. Stark Potential}

\author{Asad Ali\orcidlink{0000-0001-9243-417X}} \email{asal68826@hbku.edu.qa}
\affiliation{Qatar Center for Quantum Computing, College of Science and Engineering, Hamad Bin Khalifa University, Doha, Qatar}
\author{M.I. Hussain\orcidlink{0000-0002-6231-7746}}
\email{mahussain@hbku.edu.qa}
\affiliation{Qatar Center for Quantum Computing, College of Science and Engineering, Hamad Bin Khalifa University, Doha, Qatar}
\author{Saif Al-Kuwari\orcidlink{0000-0002-4402-7710}}
\affiliation{Qatar Center for Quantum Computing, College of Science and Engineering, Hamad Bin Khalifa University, Doha, Qatar}
\author{M. T. Rahim\orcidlink{0000-0003-1529-928X}} 
\affiliation{Qatar Center for Quantum Computing, College of Science and Engineering, Hamad Bin Khalifa University, Doha, Qatar}

\author{H. Kuniyil\orcidlink{0000-0003-0338-1278}} 
\affiliation{Qatar Center for Quantum Computing, College of Science and Engineering, Hamad Bin Khalifa University,  Doha, Qatar}
\author{Seyed~Mohammad~Hosseiny\orcidlink{0000-0002-5054-9355
}} 
\affiliation{Physics Department, Faculty of Science, Vali-e-Asr University of Rafsanjan, Rafsanjan, Iran}

\author{Jamileh Seyed-Yazdi\orcidlink{0000-0002-7345-2142}} 
\affiliation{Physics Department, Faculty of Science, Vali-e-Asr University of Rafsanjan, Rafsanjan, Iran}
\author{Hamid Arian Zad\orcidlink{0000-0002-1348-1777}}  \affiliation{Department of Theoretical Physics and Astrophysics, Faculty of Science of P. J. \v{S}af{\'a}rik University, Park Angelinum 9, 040 01 Ko\v{s}ice, Slovak Republic}
\author{Saeed Haddadi\orcidlink{0000-0002-1596-0763}} 
\email{haddadi@ipm.ir}
\affiliation{School of Particles and Accelerators, Institute for Research in Fundamental Sciences (IPM), P.O. Box 19395-5531, Tehran, Iran}

\date{\today}% It is always \today, today,
\def\be{\begin{equation}}
  \def\ee{\end{equation}}
\def\bea{\begin{eqnarray}}
\def\eea{\end{eqnarray}}
\def\f{\frac}
\def\n{\nonumber}
\def\l{\label}
\def\p{\phi}
\def\o{\over}
\def\R{\hat{\rho}}
\def\pa{\partial}
\def\om{\omega}
\def\na{\nabla}
\def\P{$\Phi$}
\begin{abstract}

% We %present a 
% numerically %exact %investigation
% investigate the one-dimensional Bose-Hubbard model under the competing influence of thermal fluctuations, a stark potential, and quenched disorder. Their collective impact on quantum coherence and phase transitions has been studied. Using full diagonalization, we compute the condensate fraction, superfluid fraction, visibility, number fluctuations, and the $\ell_1$-norm of quantum coherence to track the evolution from Mott insulator to superfluid phases. In the clean system, GS observables reveal a sharp quantum phase transition at $\tau/U \approx 0.17$, which at finite temperature transforms into a smooth crossover, suppressing long-range coherence. A stark potential ($g \neq 0$) delays this transition, localizing particles into non-SF condensates, while thermal effects activate unexpected coherence pathways via %resonant
% thermal fluctuations-induced tunneling. Disorder ($\delta = 1.0$) destroys superfluidity but preserves local coherence, with thermal states exhibiting enhanced $\ell_1$-norm of coherence in the BG regime—a signature of excited-state delocalization invisible to conventional metrics. Our results demonstrate how disorder, tilt, and temperature collaboratively reshape the coherence landscape, offering new insights for quantum simulation and the diagnosis of strongly correlated phases beyond GS physics.

We study quantum coherence and phase transitions in a finite one-dimensional Bose-Hubbard model using exact diagonalization under thermal fluctuations, a Stark potential, and disorder. We compute the condensate fraction, superfluid fraction, visibility, number fluctuations, and the $\ell_1$-norm of coherence to characterize the Mott insulator–superfluid transition. While finite-size effects prevent a sharp transition, ground-state properties reveal signatures of quantum criticality. Thermal fluctuations can enhance coherence via tunneling, a Stark potential promotes localization, and disorder suppresses global superfluidity while preserving local coherence. Our results highlight how disorder, tilt, and temperature reshape coherence and offer insights for quantum simulation and strongly correlated phases.  For systems up to six sites with unit filling, we also perform a spectral analysis through the metric mean gap ratio (MGR). However, limited statistics due to the small system size and computational constraints prevent a complete characterization of quantum chaos, yielding only approximate signatures.

\end{abstract}

% \keywords{Bose-Hubbard model, SP, quantum coherence, phase transitions}
\maketitle

\section{Introduction}  
The Bose-Hubbard model (BHM) has emerged as a cornerstone for understanding quantum phase transitions (QPTs) among cold atoms in optical lattice \cite{krutitsky2016ultracold,bloch2005ultracold,bloch2008many,fisher1989boson,Greiner2002,bloch2005ultracold,kabiraj2025}. At zero temperature, it generally exhibits a transition from superfluid (SF) to Mott insulator (MI), driven by the competition between tunnel energy and site repulsion \cite{fisher1989boson}. This transition, first experimentally observed in ultracold atoms in optical lattices \cite{Greiner2002}, highlights the delicate balance between quantum fluctuations and interactions. However, real-world implementations, whether in ultra-cold gases \cite{bloch2005ultracold,kuhner1998phases}, superconducting circuits \cite{yanay2020two,bernon2013manipulation}, or engineered quantum materials \cite{baier2016extended}, must deal with inevitable perturbations: thermal noise, quenched disorder, and Stark potential (SP). Understanding how these factors collectively reshape phase diagram of the BHM is critical for advancing both fundamental many-body physics and the development of robust quantum simulators. In a recent preprint \cite{Clavero2025}, P. M. Clavero and A. Rodríguez identified a chaotic regime in the BHM and mapped the location and extent of this chaotic phase as functions of energy and model parameters. Notably, they found that the chaotic phase of the clean BHM can be enhanced by introducing a moderate tilt. Building on this insight, we investigate the interplay between disorder strength, hopping amplitude, and SP to explore signatures of chaos in the finite-size one-dimensional (1D) disordered BHM~\cite{PhysRevLett.119.073002,PhysRevA.95.053624}.

Disorder~\cite{fisher1989boson,kisker1997bose,scalettar1991localization,bloch2008many,russ2025} and SP~\cite{schulz2019stark,yao2020many}, both break translational invariance symmetry in the BHM, represent distinct pathways to localization. Quenched disorder, arising from lattice imperfections or inhomogeneous potentials, fragments the system into a Bose glass (BG)—a compressible insulator with rare SF puddles. In contrast, SP, induced by tilting the lattice, suppresses tunneling via Wannier-Stark localization \cite{emin1987existence}, confining bosons to spatially restricted regions. While both mechanisms suppress long-range coherence, their interplay remains poorly understood, as the BHM is generally non-integrable and exhibits chaos in the presence of such potentials. 

The experimental ubiquity of these perturbations underscores their theoretical relevance. Optical lattices with laser speckle disorder \cite{Lugan2007}, tilted potentials \cite{Preiss2015}, and finite entropy densities \cite{Choi2016} demand models that transcend the pristine BHM. Traditional metrics like the condensate fraction ($f_c$) and SF density, however, often fail to resolve subtle coherence structures in such systems. For instance, the BG phase retains local coherence, invisible to global superfluidity measures, while Stark-localized states exhibit correlations absent in momentum-space visibility. Recent advances in quantum resource theory—particularly the $\ell_1$-norm of coherence \cite{Baumgratz2014}—offer a promising lens to quantify these hidden correlations, yet their application to thermally excited or tilted BHMs remains nascent. 

In this work, we systematically revisit the 1D BHM under simultaneous thermal fluctuations, SPs, and quenched disorder. By leveraging numerically exact diagonalization (ED), we dissect how these perturbations collaboratively reshape the landscape of quantum transport and coherence at zero and finite temperatures, with a focus on regimes where traditional metrics falter.   

The structure of this paper is as follows. In Section~\ref{sec:model}, we introduce the BHM with SPs and disorder, outlining the Hamiltonian and our numerical approach. Section~\ref{sec:obs} details the quantum observables and measurement metrics employed to characterize the various phases, including $f_c$, $f_s$ density, visibility ($\mathcal{V}$), $\ell_1$-norm of coherence $\mathcal{C}$, and number fluctuations. Section~\ref{sec:discussion} presents our numerical results, systematically examining the clean BHM, effects of SPs, role of disorder, and their interplay with thermal fluctuations. Finally, Section~\ref{sec:conclusion} summarizes our findings and discusses their implications.

%%%%%%%%%%%%%%%%%%%%%%%%%%%%%%%%%%%%%%%%%%%%%%%%%%%
\section{The Model}\label{sec:model}
We investigate the ground state (GS) and thermal state (TS) properties of interacting bosons confined to a 1D optical lattice within the framework of the BHM. The BHM extends the tight-binding approximation by incorporating local on-site interactions and site-resolved potentials, providing a minimal yet powerful framework for studying QPTs, disorder-induced localization, and thermal decoherence in ultracold atomic systems.

The starting point is the tight-binding Hamiltonian (TBH), which governs boson hopping between adjacent sites:
\begin{equation}
\hat{{H}}_{\mathrm{TB}} = -\frac{\tau}{2} \sum_{j=1}^{L-1} \left( \hat{a}_j^\dagger \hat{a}_{j+1} + \hat{a}_{j+1}^\dagger \hat{a}_j \right),
\label{eq:tight_binding}
\end{equation}
where $\hat{a}_j^\dagger$ ($\hat{a}_j$) creates (annihilates) spinless bosons at site $j$, with $[\hat{a}_j, \hat{a}_{j'}^\dagger] = \delta_{jj'}$, and $\tau$ denotes the tunneling amplitude. 
% In the single-particle limit, localized states $|j\rangle$ and Bloch eigenstates $|\boldsymbol{q}\rangle$ are related via
% \begin{equation}
% |\boldsymbol{q}\rangle = \frac{1}{\sqrt{L}} \sum_{j=1}^L e^{\text{i} \boldsymbol{q} \cdot r_j} |j\rangle,
% \end{equation}
% with lattice spacing $a$ and momentum $\boldsymbol{q} \in [-\pi/a, \pi/a]$. The resulting cosine dispersion relation,
% \begin{equation}
% \hat{{H}}_{\mathrm{TB}} |\boldsymbol{q}\rangle = -2\tau \cos(\boldsymbol{q}a) |\boldsymbol{q}\rangle,
% \end{equation}
% has a bandwidth of $E_{\mathrm{BW}} = 4\tau$.
Interactions are introduced via an on-site repulsion term $U$, yielding the clean BHM:
\begin{equation}
\begin{aligned}
\hat{{H}}_{\mathrm{BH}} ={} & -\frac{\tau}{2} \sum_{j=1}^{L-1} \left( \hat{a}_j^\dagger \hat{a}_{j+1} + \hat{a}_{j+1}^\dagger \hat{a}_j \right) \\
& - \mu \sum_{j=1}^{L} \hat{n}_{j} + \frac{U}{2} \sum_{j=1}^{L} \hat{n}_j(\hat{n}_j - 1),
\end{aligned}
\label{eq:BHM}
\end{equation}
where $\hat{n}_j = \hat{a}_j^\dagger \hat{a}_j$ is the number
operator. 
%The competition between $\tau$ and $U$ drives a QPT: for $\tau/U \gg 1$, delocalized SF coherence emerges ($|\psi_{\mathrm{SF}}\rangle \propto \prod_j |\alpha\rangle_j$), while $\tau/U \ll 1$ favors localized MI states ($|\psi_{\mathrm{MI}}\rangle = \prod_j |n\rangle_j$).
The competition between the tunneling amplitude $\tau$ and the on-site interaction $U$ drives a QPT in the system. In the regime $\tau/U \gg 1$, the ground state exhibits delocalized SF coherence, well approximated by a product of coherent states, $|\psi_{\mathrm{SF}}\rangle \propto \prod_j |\alpha\rangle_j$. Conversely, for $\tau/U \ll 1$, strong interactions localize the particles, favoring a MI state with fixed integer occupation at each site, $|\psi_{\mathrm{MI}}\rangle = \prod_j |n\rangle_j$.

The random site-dependent disorder $\delta_j$ is distributed within the range $[-\delta,+\delta]$ with $\delta$ being the energy scale of the strength of the disorder \cite{batrouni1992world}.
The total Hamiltonian of the clean BHM, upon inclusion of disorder and the Stark potential $gj$, is extended to:
\begin{equation}
\begin{aligned}
\hat{{H}} ={} & \hat{{H}}_{\mathrm{BH}} + \sum_{j=1}^{L} \left( g j + \delta_j\right) \hat{n}_j.
\end{aligned}
\label{eq:full_H}
\end{equation}

We solve $\hat{{H}}$ via ED in the Fock basis,  enforcing particle number conservation $\sum_j n_j = N$. The Hilbert space dimension grows combinatorially as $\mathcal{D} = \binom{L+N-1}{N}$, limiting full ED method to small system sizes \cite{zhang2010exact}.
%(typically $L \leq 8$ and $N \leq 8$). 
For $L = N = 6$, the Hilbert space contains $\mathcal{D} = 462$ states, which remains tractable.
% However, larger systems (e.g., $L = N = 12$, $\mathcal{D} > 10^6$) require approximate methods such as the density matrix renormalization group.
While finite-size effects are inherent, 1D systems typically exhibit faster qualitative convergence than higher-dimensional counterparts, due to strong quantum fluctuations that suppress long-range order. 

\begin{figure}[t]
    \centering
    \includegraphics[width=1\linewidth]{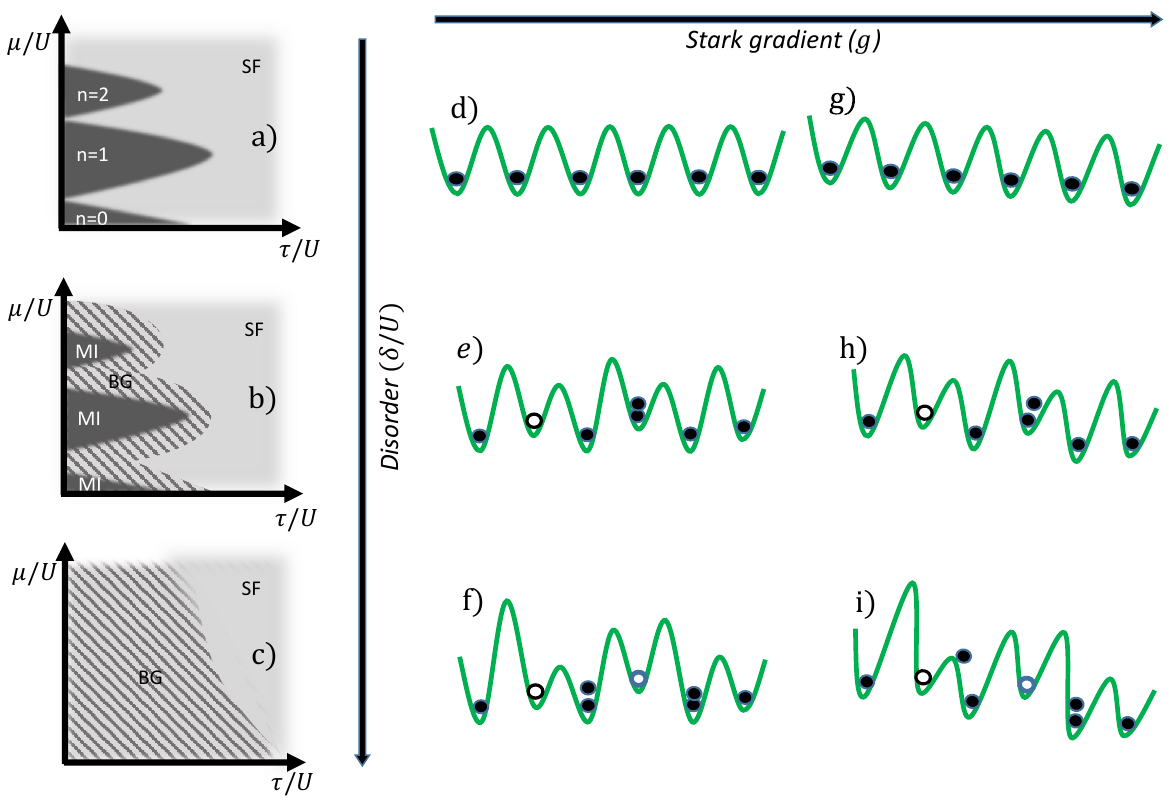}
     \caption{(a)-(c) Phase diagrams in the disordered (dirty) BHM are scaled with the system's linear site-dependent random energy offset strength $\delta/U$. (d)-(i) Single-site occupancy is shown in both the absence and presence of this offset. In the limit of negligible tunneling between lattice sites, distinct SF and MI phases emerge. As the offset becomes significant, bosonic tunneling and hopping disorder give rise to localization effects. These effects become prominent when the interaction strength satisfies $g/U>1$.}
    \label{figure1}
\end{figure}

% Figure~\ref{figure1} summarizes key phases: disorder nucleates BG regions (Figs. ~\ref{figure1} (b) and (c), while Stark fields localize particles even in the absence of interactions (Figs. ~\ref{figure1} (d)-(i)). The thermal state defined as $\hat{\varrho}_T = e^{-\beta \hat{{H}}}/\mathrm{Tr}(e^{-\beta \hat{{H}}})$, is computed from the full energy spectrum $\{|\psi_i\rangle, \mathcal{E}_i\}$, with Boltzmann weights $p_i \propto e^{-\beta \mathcal{E}_i}$. Despite truncation to $L=6$, critical features such as coherence revival persist across system sizes.
Figure~\ref{figure1} summarizes the key emergent phases: disorder BG regions [Figs.~\ref{figure1}(b) and (c)], while the presence of a SP can localize particles even in the absence of interactions [Figs.~\ref{figure1}(d)--(i)]. The thermal state, defined as $\hat{\varrho}_T = e^{-\beta \hat{H}} / \mathrm{Tr}(e^{-\beta \hat{H}})$, is obtained from the full energy spectrum $\{|\psi_i\rangle, \mathcal{E}_i\}$, with each state weighted by a Boltzmann factor $p_i \propto e^{-\beta \mathcal{E}_i}$. Although simulations are truncated to a system size of $L=6$, essential features such as coherence revivals remain robust across system sizes.

%%%%%%%%%%%%%%%%%%%%%%%%%%%%%%%%%%%%%%%%%%%%%%%%%%%
\section{Quantum Observables and Measurement Metrics}\label{sec:obs}
We employ a comprehensive set of complementary observables to characterize the equilibrium properties and QPTs in our model. Each metric captures different aspects of quantum coherence, correlations, and transport in many-body systems, providing a multifaceted picture of the underlying physics.

The condensate fraction $f_c$ quantifies the fraction of Bose-Einstein condensation according to the Penrose-Onsager criterion \cite{Onsager1956}. It is defined as the ratio of the largest eigenvalue $\lambda_{\text{max}}$ of the one-body density matrix 
% $\rho^{(1)}$ 
to the total particle number, namely, $f_c = {\lambda_{\text{max}}}/{N}$,
where the one-body density matrix is defined as $\rho^{(1)}_{ij} = \langle\hat{a}^\dagger_i \hat{a}_j\rangle$. For the GS, the one-body correlation function is given by \cite{zhang2010exact}  $\langle\hat{a}^\dagger_i \hat{a}_j\rangle_{\text{GS}} = \langle\psi_0|\hat{a}^\dagger_i \hat{a}_j|\psi_0\rangle$, while for TS, it is computed as $\langle\hat{a}^\dagger_i \hat{a}_j\rangle_{\text{TS}} = \mathrm{Tr}(\hat{\varrho}_T\hat{a}^\dagger_i \hat{a}_j)$. In the SF phase, $f_c\rightarrow 1$, indicating macroscopic occupation of the lowest momentum state. In contrast, $f_c$ can be significantly suppressed in the MI and BG phases.
On the other hand, in the SF, $f_s$ characterizes the emergence of superfluidity by quantifying the fraction of particles that can flow without viscosity. In the BHM, $f_s$ is computed by analyzing the system's response to a phase twist applied to the hopping terms of the TBH.

Applying a phase twist $\phi$ modifies the hopping terms in Eq. \eqref{eq:full_H} as
\begin{equation}
\hat{{H}}_{\mathrm{TB}}(\phi) = -\frac{\tau}{2} \sum_{j=1}^{L-1} \left( e^{\text{i}\phi} \hat{a}_j^\dagger \hat{a}_{j+1} + e^{-\text{i}\phi} \hat{a}_{j+1}^\dagger \hat{a}_j \right).
\label{newtbh}
\end{equation}
Substituting this new TBH \eqref{newtbh} into Eq. \eqref{eq:full_H}, the SF fraction in the GS $|\psi_0\rangle$ is determined from the second derivative of the GS energy $\mathcal{E}_{0}(\phi)$ with respect to $\phi$ at $\phi = 0$ \cite{kuhner2000one,leggett1998superfluid,buonsante2009gutzwiller,gerster2016superfluid}:
\begin{equation}
f_s(|\psi_0\rangle) = \frac{L^2}{N \tau} \left. \frac{d^2 \mathcal{E}_{0}(\phi)}{d\phi^2}\right|_{\phi=0},
\end{equation}
where $N$ is the total particle number and $L$ is the system size.

At $T>0$, the SF for a TS is similarly obtained from the free energy $F(\phi) = -\beta^{-1} \ln Z(\phi)$ \cite{rousseau601681exact,prokof2000two,lu2006finite}
with
$Z(\phi) = \text{Tr} (\exp[-\beta \hat{{H}}(\phi)] )$
as
\begin{equation}
f_s(\hat{\varrho}_T) = \frac{L^2}{N\tau} \left. \frac{d^2 F(\phi)}{d\phi^2}\right|_{\phi=0}.
\end{equation}

The second derivatives $\frac{d^2 \mathcal{E}_{0}(\phi)}{d\phi^2}$ and $\frac{d^2 F(\phi)}{d\phi^2}$ quantify the system's rigidity to phase twists, directly measuring superfluidity \cite{prokof2000two}. While $f_s(|\psi_0\rangle)$ captures purely quantum effects at zero temperature, $f_s(\hat{\varrho}_T)$ incorporates thermal fluctuations. Both quantities vary across the SF–MI QPT, showing the interplay among interactions, disorder, and thermal effects in the BHM.

The momentum distribution $n_k$ reveals coherence properties in reciprocal space and is directly accessible in cold atom experiments through time-of-flight measurements. It is defined as the Fourier transform of the one-body correlation function \cite{ejima2011dynamic,varney2008quantum,roux2008quasiperiodic}
\begin{equation}
    n_k = \frac{1}{L}\sum_{i,j=1}^L e^{\text{i}k(i-j)}\langle\hat{a}^\dagger_i \hat{a}_j\rangle,
\end{equation}
where $k = \frac{2\pi l}{L}$ represents the discrete momentum values with $l \in \{0, 1, \ldots, L-1\}$. The visibility $\mathcal{V}$ quantifies the contrast in the momentum distribution:
\begin{equation}
    \mathcal{V} = \frac{n_k^{\text{max}} - n_k^{\text{min}}}{n_k^{\text{max}} + n_k^{\text{min}}},
\end{equation}
where $n_k^{\text{max}}$ and $n_k^{\text{min}}$ are the maximum and minimum values of the momentum distribution. In the SF phase, $\mathcal{V}$ approaches unity due to sharp peaks at $k=0$, while in the MI or localized phases, $\mathcal{V}$ is significantly reduced.

The following formula provides a basis-dependent quantification of quantum coherence within the density matrix $\rho$, expressed in the Fock basis:
\begin{equation}
    \mathcal{C} = \sum_{i\neq j}|\rho_{ij}| = \sum_{i,j}|\rho_{ij}| - \sum_{i}|\rho_{ii}|.
\end{equation}
This resource-theoretic measure captures the overall quantum superposition present in the system. For pure states, $\mathcal{C}$ measures the degree of delocalization across Fock states, while for mixed states, it reflects both quantum coherence and contributions from coherent mixtures of excited states (which can be thermally populated).

The global fluctuation $\mathcal{F}$ characterizes the quantum uncertainty in particle number at each site \cite{plimak2004occupation}, averaged over the entire lattice:
\begin{equation}
    \mathcal{F} =\frac{1}{L} \sum_{i=1}^L \sqrt{\langle\hat{n}_i^2\rangle - \langle\hat{n}_i\rangle^2},
\end{equation}
where $\hat{n}_i = \hat{a}^\dagger_i\hat{a}_i$ is the number operator at site $i$. In the MI phase, number fluctuations are strongly suppressed ($\mathcal{F} \approx 0$), while in the SF phase, $\mathcal{F}$ increases significantly due to particle delocalization.

These five observables offer complementary insights into our model. The $f_c$ and $\mathcal{C}$ indicate quantum coherence, while the  $f_s$ reflects transport capabilities due to coherence. Coherence can persist without superfluidity, as seen in the BG phase. Spatially,  $\mathcal{V}$ and SF fraction $f_s$ act as global probes of long-range coherence and flow, while number fluctuation $\mathcal{F}$ provides local information about quantum uncertainty. Experimentally, the momentum distribution (and thus visibility) can be extracted from time-of-flight measurements, while number fluctuations are accessible through \textit{in situ} imaging. These measurements comprehensively characterize quantum phases, coherence properties, and phase transitions across varying interaction strengths, disorder, and temperatures.

% \begin{figure}
%     \centering
%     \includegraphics[width=0.9\linewidth]{f2.pdf}
%     \caption{Caption}
%     \label{fig:enter-label}
% \end{figure}

\section{Results and Discussions}\label{sec:discussion}

We examine the quantum mechanical and thermal properties of bosons in 1D BHM across various parameter regimes. First, analyzing the clean BHM to establish baseline behavior, we then investigate the impact of SPs and random disorder separately, focusing on how these perturbations modify QPTs and coherence properties.

\subsection{Clean BHM}
\begin{figure}[t]
    \centering
    \includegraphics[width=1.0\linewidth]{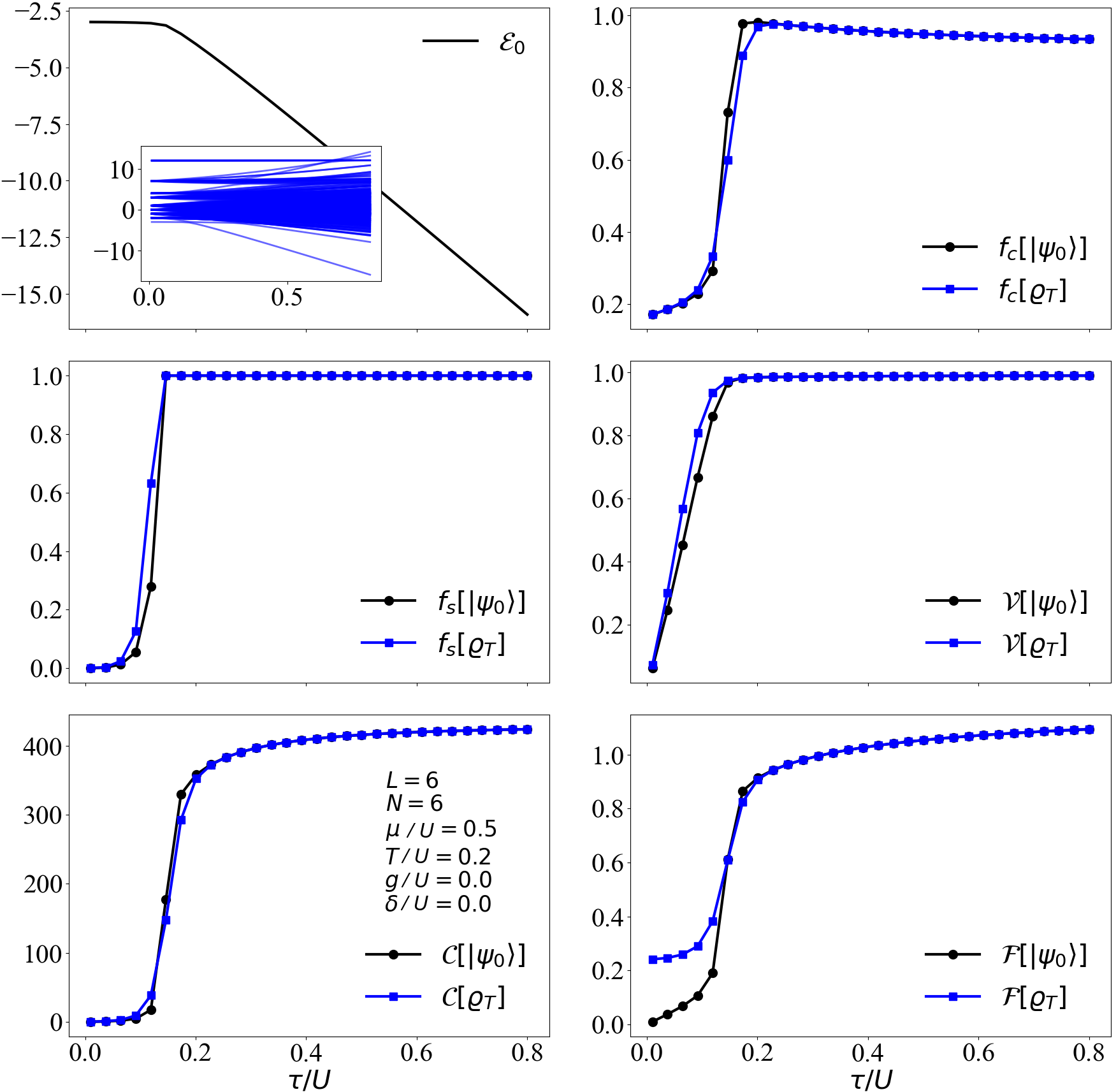}
    \put(-225,220){\textbf{a)}}
    \put(-100,220){\textbf{b)}}
    \put(-225,150){\textbf{c)}}
    \put(-103,150){\textbf{d)}}
    \put(-225,68){\textbf{e)}}
    \put(-100,63){\textbf{f)}}
        % \put(-210,230){\rotatebox{180}{$\uparrow$}}
        % \put(-205,183){\rotatebox{0}{$\uparrow$}}
        % \put(-210,150){\rotatebox{180}{$\uparrow$}}
        % \put(-210,36){\rotatebox{180}{$\uparrow$}}
        % \put(-84,227){\rotatebox{180}{$\uparrow$}}
        % \put(-84,150){\rotatebox{180}{$\uparrow$}}        \put(-84,60){\rotatebox{180}{$\uparrow$}}
        
\caption{
Clean BHM ($g/U = 0$, $\delta/U = 0$) observables at unit filling with $\mu/U = 0.5$ versus $\tau/U$ for the both GS $|\psi_0\rangle$ (black circles) and TS $\varrho_T$ with $T/U = 0.2$ (blue squares), assuming $L=N=6$. (a) Ground state energy $\mathcal{E}_0$ with energy spectrum (the inset) showing gap closure near the transition point $\tau/U \approx 0.17$, which is acceptable for the given system size \cite{kuhner1998phases,satoshi2012characterization,carrasquilla2013scaling,kiely2022superfluidity} under open boundary conditions. (b) Condensate fraction $f_c$; (c) SF fraction $f_s$, (d) visibility $\mathcal{V}$, (e) $\ell_1$-norm of coherence $\mathcal{C}$, and (f) number fluctuations $\mathcal{F}$.
}
    \label{figure2}
\end{figure}

Figure~\ref{figure2} shows key observables versus $\tau/U$ for the clean BHM ($\delta/U = g/U = 0$), comparing GS (black circles) with TS properties ($T/U = 0.2$, blue squares). The GS energy $\mathcal{E}_0$ decreases monotonically with increasing $\tau/U$ as kinetic contributions grow (Fig.~\ref{figure2}(a)). The many-body spectrum shown in the inset reveals a gap closing at $\tau/U \approx 0.17$ that hints at the SF-MI QPT.

\begin{figure}
\centering
\includegraphics[width=1\columnwidth]{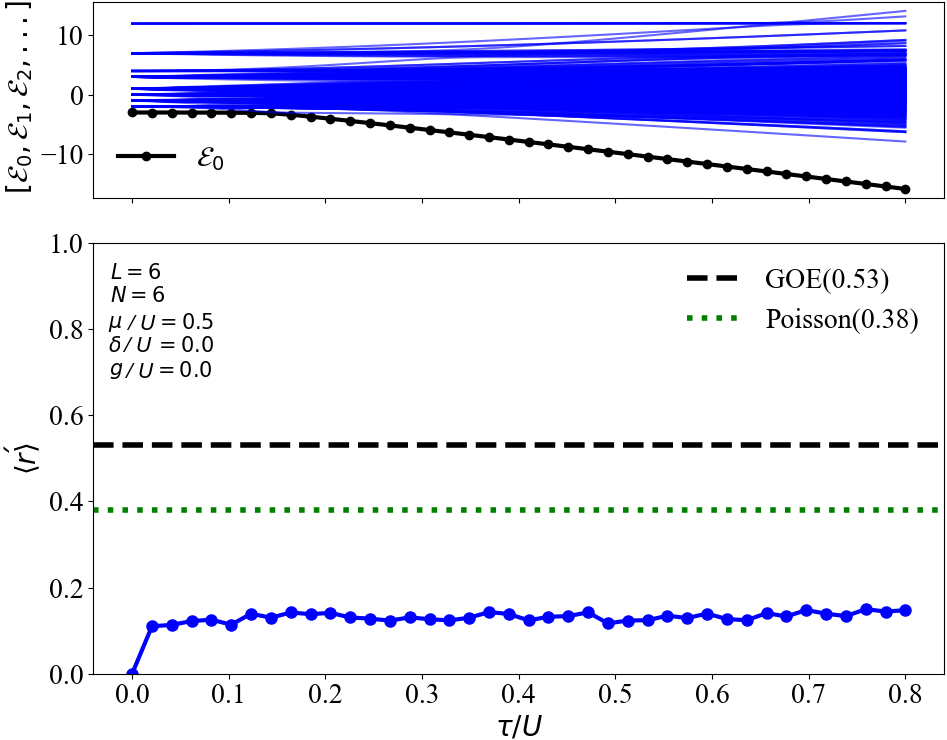}
\caption{%
Many-body spectral properties of the clean BHM with $L=N=6$ and $\mu/U = 0.5$ at unit filling as a function of normalized hopping $\tau/U$. \textbf{(top)} Eigenenergies $\mathcal{E}_n(\tau/U)$: GS energy $\mathcal{E}_0$ decreases and the excited-state band broadens, with a gap closure at $\tau/U \approx 0.17$, indicating the MI–SF transition. \textbf{(bottom)}  $\langle r'\rangle \approx 0.15$ remains constant, indicating persistent spectral correlations.
% and high symmetry.
}
\label{fig:energy_vs_tau}
\end{figure}

In the strong onsite-interaction regime ($\tau/U < 0.17$), all metrics including $f_c$ (Fig.~\ref{figure2}(b)), $f_s$ (Fig.~\ref{figure2}(c)), visibility $\mathcal{V}$ (Fig.~\ref{figure2}(d)), and $\ell_1$-norm of coherence $\mathcal{C}$ (Fig.~\ref{figure2}(e)) remain suppressed, confirming the localized nature of the MI phase. Notice that the overall number fluctuations $\mathcal{F}$ (Fig.~\ref{figure2}(f)) are minimal, demonstrating the characteristic incompressibility of this phase.

Beyond $\tau/U \approx 0.17$, all observables sharply increase, signaling the emergence of global phase coherence and particle delocalization in the SF phase. The QPT appears sharp for GS but is significantly broadened for TS due to thermal fluctuations. The TS exhibits systematically lower $f_c$ and $f_s$ values, highlighting thermal disruption of long-range coherence, with effects most pronounced in the weakly interacting regime ($\tau/U \gtrsim 0.3$).

Interestingly, in the intermediate regime ($0.1 < \tau/U < 0.2$), the TS displays slightly enhanced $\mathcal{C}$ and $\mathcal{F}$ compared to GS. This seemingly counterintuitive behavior stems from thermal occupation of low-lying excited states with greater spatial delocalization, which increases local coherence and number fluctuations despite reducing global order.

To characterize the statistical properties of the energy spectrum, we compute the mean gap ratio (MGR) of consecutive energy level spacings $\langle r' \rangle$, defined as:
\begin{equation}
  \langle r'\rangle
  = \Big\langle \frac{\min(s_n,s_{n+1})}{\max(s_n,s_{n+1})} \Big\rangle,
  \quad s_n = \mathcal E_{n+1}-\mathcal E_n.
\end{equation}
where $s_n$ represents the spacing between adjacent energy levels. The MGR is a robust metric for distinguishing between spectral statistics without requiring unfolding procedures. For systems exhibiting quantum chaos and localization, the energy levels follow Gaussian Orthogonal Ensemble (GOE) statistics with a theoretical MGR of $\langle r' \rangle \approx 0.53$. In contrast, integrable systems typically follow Poisson statistics with $\langle r' \rangle \approx 0.38$. 

The level‐spacing analysis of the clean BH chain shows that even as the hopping $\tau/U$ increases from zero, where the system is an MI, the MGR quickly settles to a constant value $\langle r' \rangle \approx 0.15$ (see Fig.~\ref{fig:energy_vs_tau}). This value lies well below the Poisson and GOE benchmarks, implying that the spectrum retains non‐ergodic correlations. Such persistent level correlations suggest the presence of approximate conservation laws or hidden symmetries that survive across the MI‐SF crossover.

\begin{figure}
    \centering
    \includegraphics[width=1\linewidth]{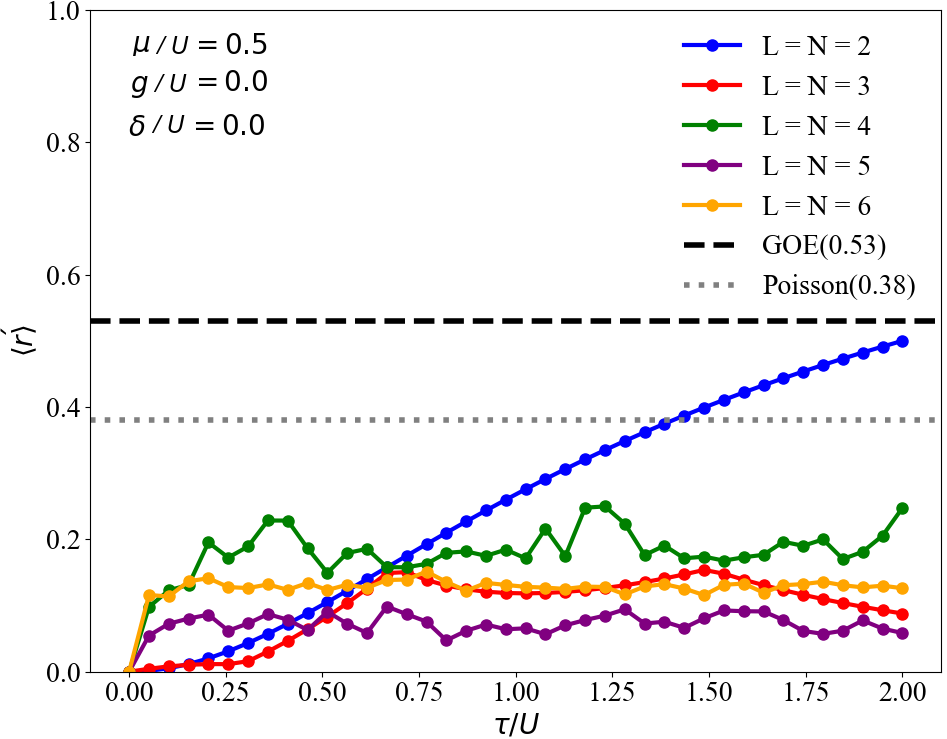}
\caption{System-size dependence of the $\langle r' \rangle$ versus  $\tau/U$ for the clean BHM ($g/U = 0.0$, $\mu/U = 0.5$, $\delta/U = 0.0$) at unit filling. Different system sizes ($L = N = 2$ through $L = N = 6$) are shown with their respective colored curves. The dashed and dotted horizontal lines represent the theoretical values for GOE (0.53) and Poisson (0.38) statistics, respectively. While the smallest system ($L = N = 2$) shows increasing $\langle r' \rangle$ with $\tau/U$ that approaches the GOE limit, larger systems maintain consistently low $\langle r' \rangle$ values (0.05-0.25) across all hopping strengths, demonstrating persistent non-ergodic spectral correlations that remain stable through the SF-MI transition at $\tau/U \approx 0.17$. For the minimal two-site case ($L=N=2$), the apparent GOE-like value of the mean gap ratio
$\langle r' \rangle$ is a finite-size artifact rather than genuine chaos. With only three
many-body states, the model is integrable, but poor spectral statistics, boundary effects, and
kinetic-energy dominance mimic level repulsion. For larger systems ($L \geq 3$), these artifacts
disappear and $\langle r' \rangle$ settles at sub-Poissonian values, consistent with nonergodic
behavior.
}
    \label{f4}
\end{figure}

Figure~\ref{f4} extends our analysis of clean BHM by examining MGR across system sizes from $L = N = 2$ to $L = N = 6$, all at unit filling. Although Fig.~\ref{figure2} shows the MGR saturates around $\langle r' \rangle \approx 0.15$ for a fixed system size, this also highlights the critical role of the system size in shaping spectral statistics.

For the smallest interacting chain ($L = N = 2$, blue curve), the MGR increases monotonically with $\tau/U$, eventually approaching the GOE value of 0.53 at large hopping strengths. This indicates that quantum chaotic behavior emerges more readily in minimal systems as kinetic energy and edge effects due to finite size become dominant.

In contrast, relatively larger chains ($3\leq L = N \leq 6$) exhibit consistently low MGR values (mostly within 0.05--0.25) across all $\tau/U$, staying well below both the Poisson limit (0.38) and the GOE limit (0.53). This persistent sub-Poissonian behavior suggests that the non-ergodic spectral correlations seen in Fig.~\ref{figure2} are robust features of the clean BHM at unit filling as the system size increases. 
Interestingly, for $L = N \geq 3$, MGR shows no notable change near the SF-MI transition point ($\tau/U \approx 0.17$), despite the sharp changes in $f_c$, $f_s$, and coherence seen in Fig.~\ref{figure2}. This indicates that while conventional observables signal the transition, spectral statistics overall remain stable beyond minimal system sizes.

These results imply that approximate conservation laws or hidden symmetries underlying the spectral structure become increasingly robust with system size, preserving non-ergodic behavior even as the system transitions from localized (MI) to delocalized (SF) regimes.

\begin{figure}[t]
    \centering
    \includegraphics[width=1\linewidth]{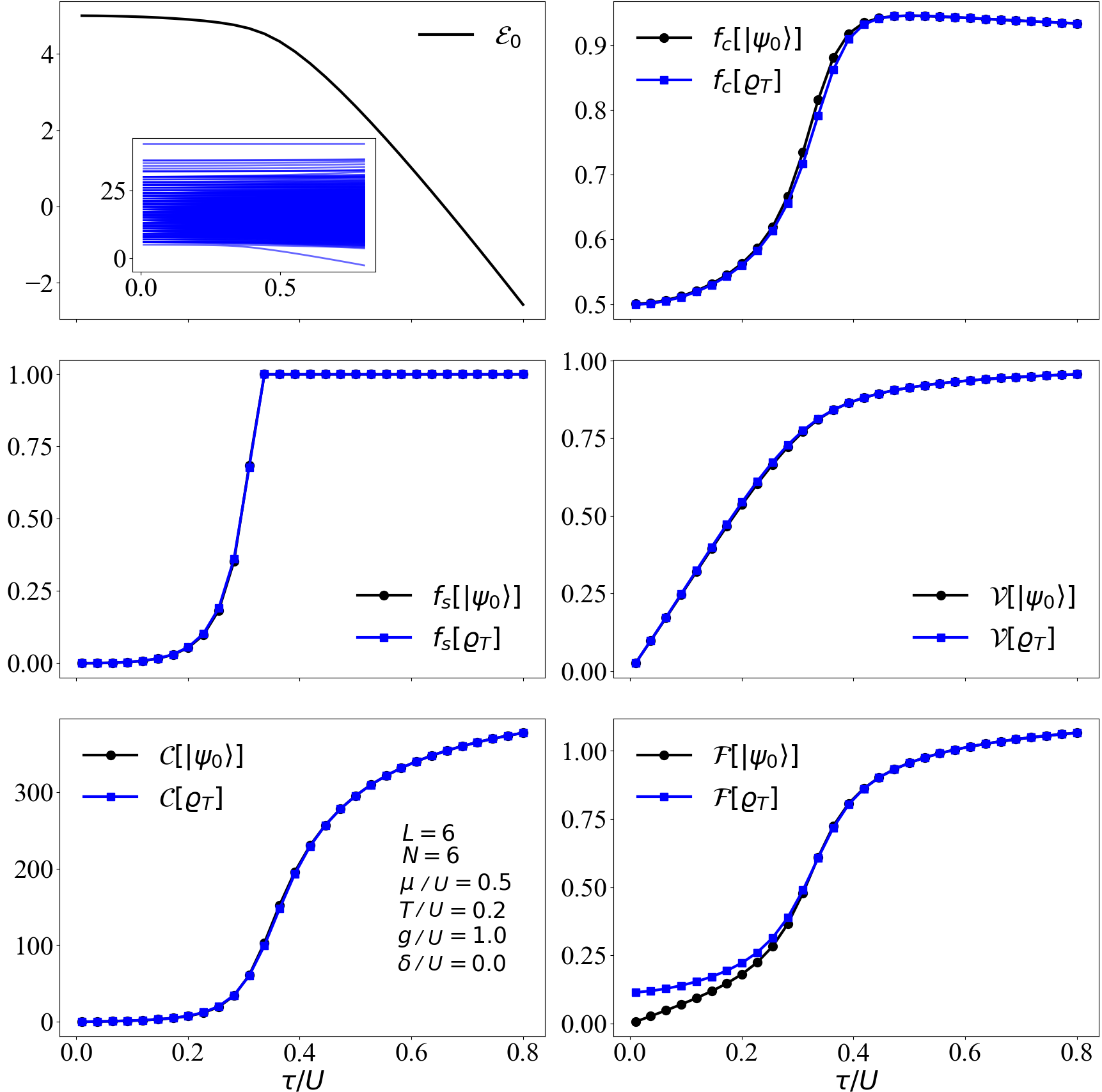}
    \put(-140,200){\textbf{a)}}
    \put(-20,200){\textbf{b)}}
    \put(-140,120){\textbf{c)}}
    \put(-20,120){\textbf{d)}}
    \put(-140,18){\textbf{e)}}
    \put(-20,18){\textbf{f)}}
\caption{
The BHM with strong SP ($g/U = 1.0$, $\delta/U = 0.0$). GS ($|\psi_0\rangle$, black circles) and TS ($\varrho_T$, blue squares) observables versus $\tau/U$ for $L = N = 6$, $\mu/U = 0.5$, and temperature $T/U = 0.2$. (a) GS energy $\mathcal{E}_0$ with spectrum inset showing Wannier-Stark localization, (b) $f_c$, (c)  $f_s$, (d) $\mathcal{V}$, (e)  $\mathcal{C}$, and (f) $\mathcal{F}$.
}   
    \label{figure3}
\end{figure}

\begin{figure}[t]
\centering
\includegraphics[width=1\columnwidth]{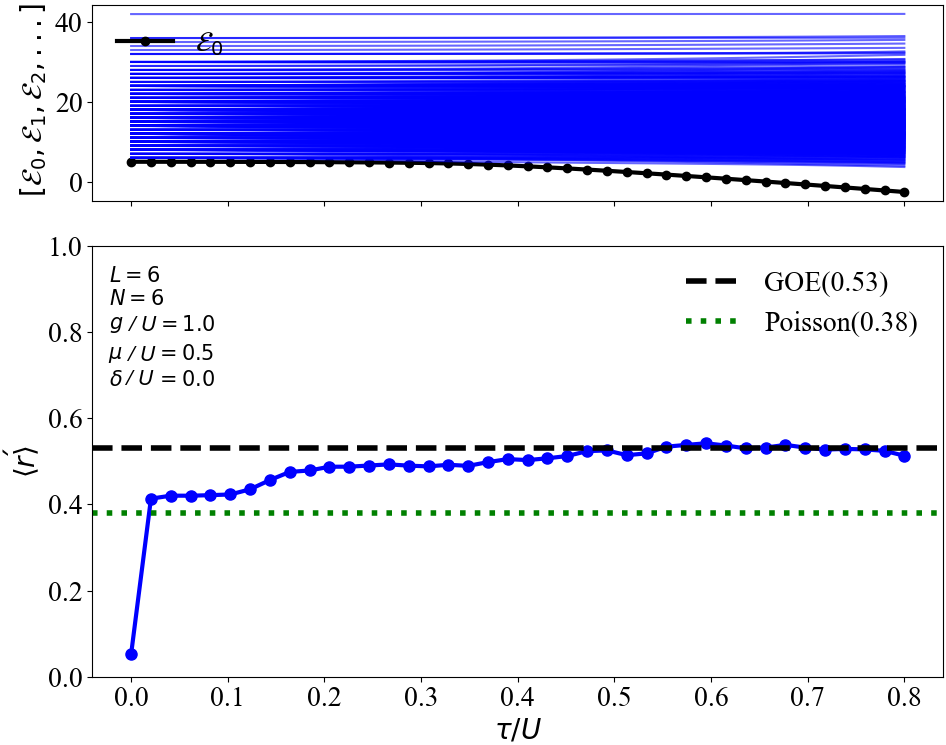}
\caption{%
Spectral statistics of the BH chain with SP for $L = N = 6$,  $g/U = 1.0$, $\mu/U = 0.5$, and $\delta/U = 0$, as a function of normalized hopping $\tau/U$. \textbf{(top)} Eigenenergies $\mathcal{E}_n(\tau/U)$ 
\textbf{(bottom)} $\langle r' \rangle$ versus $\tau/U$. Dashed lines indicate Poisson (0.38) and GOE (0.53) limits.  The SP drives a transition from localized to semi-ergodic behavior, inhibiting full thermalization.}
\label{fig:interacting_spectrum}
\end{figure}

\begin{figure}
    \centering
    \includegraphics[width=1\linewidth]{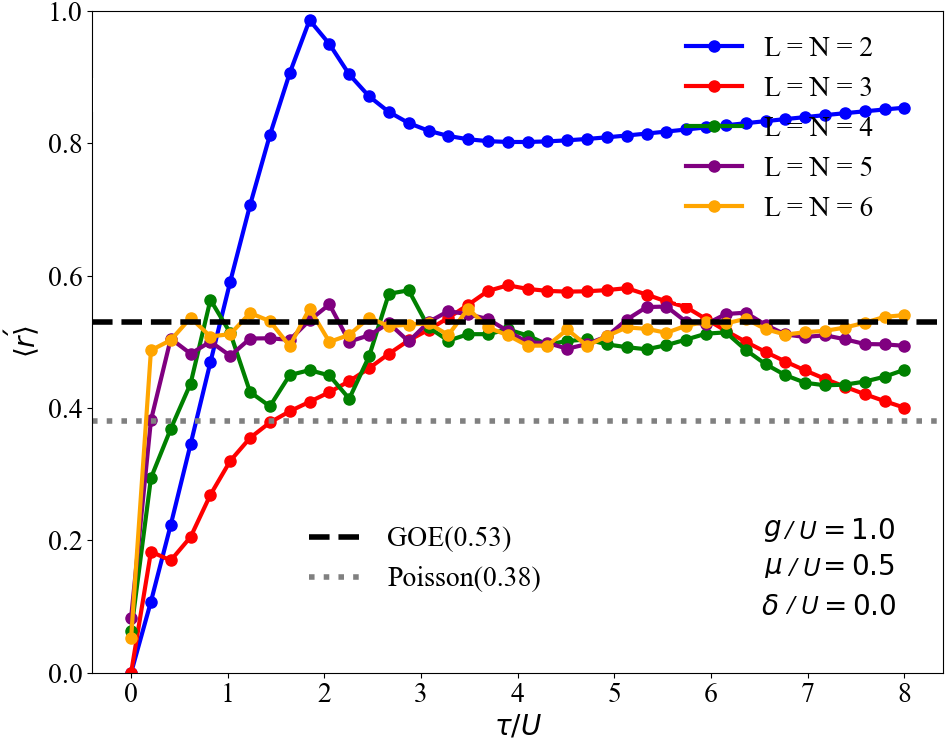}
\caption{System-size dependence of the $\langle r' \rangle$ versus $\tau/U$ for the BHM with strong SP ($g/U = 1.0$, $\mu/U = 0.5$, $\delta/U = 0.0$) at unit filling. Different system sizes ($L = N = 2$ through $L = N = 6$) are shown with their respective colored curves. The dashed and dotted horizontal lines represent the theoretical values for GOE (0.53) and Poisson (0.38) statistics. All system sizes show a transition from Poisson-like statistics at small $\tau/U$ to values approaching or exceeding the GOE limit as hopping increases. The smallest system ($L = N = 2$) exhibits a pronounced peak exceeding the GOE limit near $\tau/U \approx 2.0$ before settling at $\langle r' \rangle \approx 0.8$ at large hopping. Larger systems show oscillatory behavior around the GOE value, indicating a complex interplay between SP-induced localization and hopping-induced delocalization.}    
\label{f7}
\end{figure}

\begin{figure}[t]
    \centering
\includegraphics[width=1\linewidth]{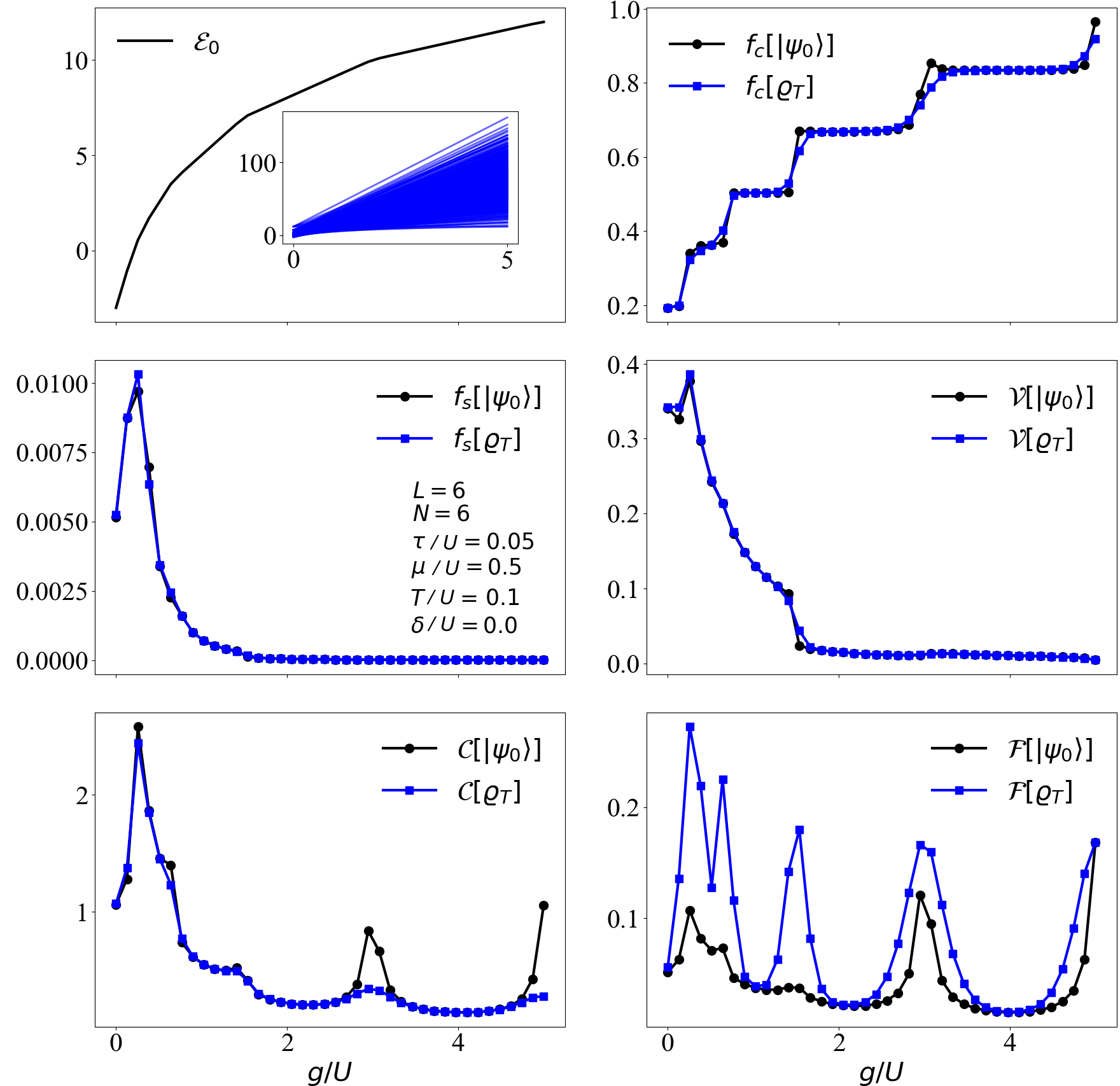}
    \put(-175,205){\textbf{a)}}
    \put(-30,205){\textbf{b)}}
    \put(-175,120){\textbf{c)}}
    \put(-30,120){\textbf{d)}}
    \put(-175,50){\textbf{e)}}
    \put(-30,50){\textbf{f)}}
\caption{SP-induced suppression in clean BHM ($\delta = 0$). GS (black circles) and TS at $T/U = 0.1$ (blue squares). %observables versus SP strength $g$. Fixed parameters: $L = N = 6$, $\tau = 0.05$, $U = 1.0$, $\mu = 0.5$. 
(a) GS energy $\mathcal{E}_0$ with spectrum inset, (b) $f_c$, (c)  $f_s$, (d) $\mathcal{V}$, (e)  $\mathcal{C}$, and (f)  $\mathcal{F}$.}
    \label{figure6}
\end{figure}

\subsection{Effects of SP}
Next, we investigate the 1D BHM with a strong SP ($g/U = 1.0$) while maintaining a clean system ($\delta/U = 0$). Figure~\ref{figure3} presents observables versus $\tau/U$ for both GS and TS at low temperature $T/U = 0.2$.
The GS energy shown in Fig.~\ref{figure3}(a) remains relatively flat for $\tau/U \lesssim 0.3$, indicating suppressed kinetic energy gain due to the effect of SP. The many-body spectrum shows no prominent sharp gap closing, unlike Fig. \ref{figure2}(a), suggesting that the SF-MI transition is smeared by breaking the translation symmetry as a result of the tilt potential, consistent with Wannier-Stark localization, where particles become energetically confined to individual sites or small clusters.

The $f_c$ (Fig.~\ref{figure3}(b)) increases steadily with $\tau/U$ but lacks the sharp crossover seen in the clean case and saturates at a lower value ($\sim$0.95 versus $\sim$0.99), indicating partial suppression of Bose-Einstein condensation. TS and GS curves closely overlap, suggesting that the tilt predominantly limits condensation rather than the given temperature.

The most revealing fact in $f_s$ (Fig.~\ref{figure3}(c)), which remains nearly zero for $\tau/U \lesssim 0.3$ before rising sharply. This delayed onset of superfluidity reflects the energetic cost of tunneling against the field SP, which must be overcome by sufficiently strong kinetic energy. The thermal curve closely follows the GS, emphasizing the primacy of the SP in shaping this QPT.

The visibility $\mathcal{V}$ (Fig.~\ref{figure3}(d)) increases more gradually than that of the clean case, with modest thermal effects. The $\ell_1$-norm of coherence $\mathcal{C}$ (Fig.~\ref{figure3}(e)) shows a delayed onset near $\tau/U \approx 0.2$, with the TS exhibiting slightly larger coherence than that of the GS throughout the transition—highlighting thermal occupation of delocalized excited states which enhance local coherence even without global phase stiffness.

Number fluctuations $\mathcal{F}$ (Fig.~\ref{figure3}(f)) remain suppressed at small $\tau/U$, confirming the incompressibility of the localized phase, and increase steadily in the SF regime. Notably, $\mathcal{F}(\varrho_T)$ exceeds $\mathcal{F}(|\psi_0\rangle)$ for $\tau/U \lesssim 0.4$, indicating enhanced thermal activation of particle-hole excitations. The results in Fig. \ref{figure3} demonstrate that the SP significantly delays and smooths the onset of coherence and superfluidity.

Figure~\ref{fig:interacting_spectrum} further clarifies the role of the SP by tracking the  $\langle r' \rangle$ as a function of $\tau/U$. At weak hopping ($\tau/U < 0.2$), the system exhibits strong localization with $\langle r' \rangle \approx 0.38$, consistent with Poisson statistics. This localized regime arises from the effects of the nonzero SP ($g/U = 1.0$), while the chemical potential ($\mu/U = 0.5$) modulates the level structure without restoring the delocalization under the unit filling condition.
In the crossover regime ($0.2 < \tau/U < 0.6$), increasing hopping begins to overcome localization, and $\langle r' \rangle$ increases. However, even at large hopping ($\tau/U > 0.6$), $\langle r' \rangle$ saturates below the GOE value (0.53), reaching only $\approx 0.48$. This indicates partial ergodicity: the system thermalizes only incompletely.
These results show that the SP reshapes the ergodic landscape by stabilizing a semi-ergodic regime, where residual localization persists despite strong kinetic energy.

% Compared to untilted systems, the presence of $g = 1.0$ both extends the localized phase and limits full thermalization.

Figure~\ref{f7} examines the  $\langle r' \rangle$ versus  $\tau/U$ for the BHM with $g/U = 1.0$ across various system sizes at unit filling case. This analysis reveals fundamental changes in spectral statistics induced by the competition between SP and hopping.
At small $\tau/U$, all system sizes exhibit $\langle r' \rangle$ values near zero, indicating strong spectral correlations consistent with localized states dominated by the SP. As $\tau/U$ increases, we observe a rapid transition toward Poisson statistics ($\langle r' \rangle \approx 0.38$) and beyond, with system size significantly influencing this evolution.

The smallest system ($L = N=2$, blue curve) shows the most dramatic behavior, with $\langle r' \rangle$ overshooting the GOE value (0.53) to reach a pronounced peak of approximately 0.98 near $\tau/U \approx 2.0$ before stabilizing around 0.85 at larger hopping strengths. This super-GOE behavior indicates an exceptionally strong level of repulsion in this minimal system due to the extremely small size effect.
The systems with $L = N=3$ and $L = N=4$ shown by red and green curves display non-monotonic behavior. Namely, with an increase of $\tau/U$, $\langle r' \rangle$ initially exceeds above the Poisson value, successively exhibiting oscillations in the range 0.4-0.6, and ultimately declines toward the Poisson limit at the high values of $\tau/U$. This suggests a complex progression through different regimes of spectral statistics as hopping competes with the SP.

The larger systems ($L = N=5,6$, purple and orange curves) show the most rapid initial increase in $\langle r' \rangle$ with $\tau/U$, reaching the GOE value at relatively small $\tau/U \approx 0.5$. For intermediate and large hopping strengths, these systems stabilize with $\langle r' \rangle$ values fluctuating around the GOE benchmark, indicating substantial but incomplete ergodicity.
These results contrast sharply with our observations for the clean BHM (Fig.~\ref{f4}), where larger systems maintained consistently low $\langle r' \rangle$ values (0.05-0.25) across all hopping strengths. The introduction of the SP fundamentally alters the spectral landscape, enabling a transition toward quantum chaotic statistics at sufficient hopping strengths.

The progression from near-zero to Poisson and approaching GOE statistics with increasing $\tau/U$ reflects the competition between the localizing influence of the SP and the delocalizing effect of tunneling. The system-size dependence suggests that finite-size effects significantly impact this competition, with the smallest system exhibiting the most extreme deviation from GOE statistics at intermediate hopping strengths.

Now, we analyze the effects of varying SP strength on coherence properties. Figure~\ref{figure6} shows observables versus SP strength $g/U \in [0, 5]$ at fixed $\tau/U = 0.05$, and $T/U = 0.1$ in a clean system.
The GS energy shown in Fig.~\ref{figure6}(a) increases monotonically with $g/U$, reflecting the growing cost of occupying sites away from the potential minimum. The many-body spectrum exhibits increasing level crowding and band flattening, characteristic of Wannier-Stark localization.
Figure ~\ref{figure6}(b) shows striking non-monotonic behavior of $f_c$ with respect to $g/U$. It remains low at small $g/U$ due to suppressed tunneling but increases in steps as localized single-particle states hybridize. At large $g/U \gtrsim 4$, $f_c$ sharply rises toward unity, signaling condensate localization into the lowest-energy site. Thermal effects primarily smooth these transitions.

The SF fraction $f_s$ versus $g/U$ is illustrated in Fig.~\ref{figure6}(c). This quantity for both GS and TS exhibits a single peak near $g/U \approx 0.2$ and rapidly drops to zero beyond $g/U \gtrsim 1$. This indicates the extreme sensitivity of phase rigidity to potential SPs, such that even moderate tilts can disrupt global phase coherence. The collective nature of $f_s$ is essentially disrupted by SP-induced localization, as evidenced by the minimal thermal effects. %The thermal effects on $f_s$ are minimal, indicating its collective nature is fundamentally disrupted by SP-induced localization.

The visibility $\mathcal{V}$ (Fig.~\ref{figure6}(d)) follows a similar trend, peaking at small $g/U$ and decreasing at larger values due to loss of long-range interference. The TS visibility remains consistently below the GS due to thermal decoherence.

Besides, the $\ell_1$-norm of coherence $\mathcal{C}$ (Fig.~\ref{figure6}(e)) exhibits oscillations on a decaying background. These fluctuations result from resonances between neighboring sites at specific $g/U$ values where inter-site energy differences match tunneling strengths. At these points, coherence temporarily increases despite strong SPs. The TS shows more pronounced coherence peaks, suggesting a thermal population of these resonant states.

Finally, the number fluctuations $\mathcal{F}$ (Fig.~\ref{figure6}(f)) reflect the interplay of interactions, tilt, and thermal effects. They decline with increasing SP due to particle localization, while TS maintains larger $\mathcal{F}$ over a broader range, particularly near resonant $g/U$ values where delocalization reemerges.
These results demonstrate that while linear SPs generally induce strong localization and suppress SF and coherence, both GS and TS observables exhibit complex and non-monotonic behaviors with resonance-enhanced coherence even in the strongly localized regime.

\begin{figure}[t]
\centering
\includegraphics[width=1\columnwidth]{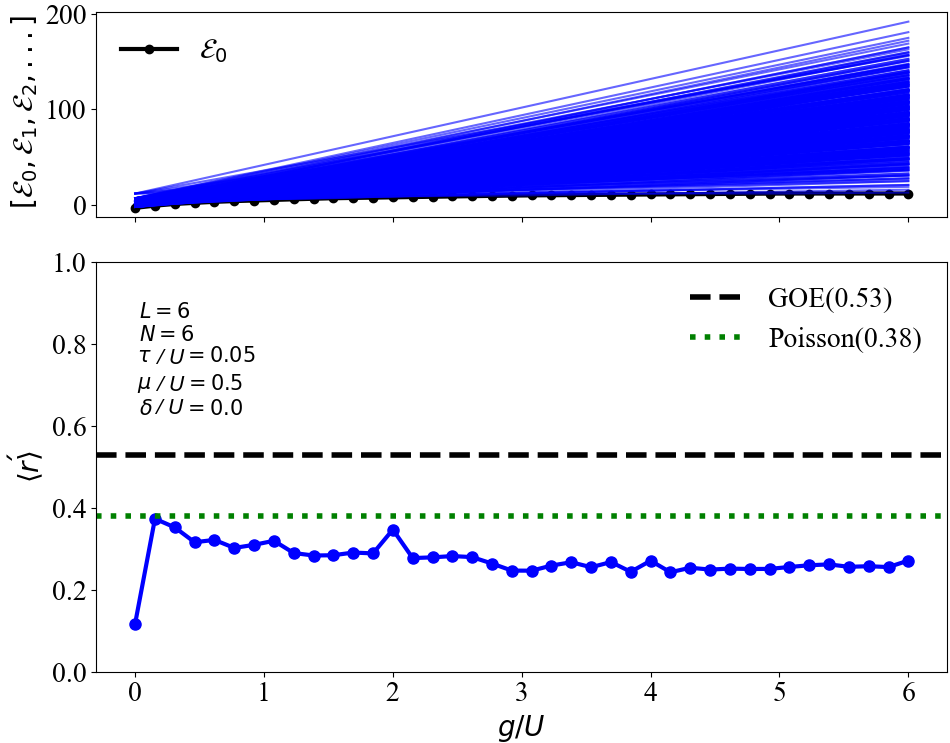}
\caption{Evolution of $\langle r' \rangle$ with increasing interaction strength $g/U$ for fixed hopping $\tau/U=0.05$. Horizontal lines mark the GOE (0.53) and Poisson (0.38) limits. System parameters: $L=6$, $N=6$, $\tau/U=0.05$, $\mu/U=0.5$, $\delta/U=0.0$.}
\label{f9}
\end{figure}

Top panel of Fig. \ref{f9} presents the eigenenergy spectrum and MGR as functions of the SP strength $g/U$ for a system with $L=6$ sites, $N=6$ bosons,  negligibly small tunneling amplitude $\tau/U=0.05$, chemical potential $\mu/U=0.5$, and zero disorder ($\delta/U=0.0$).
We observe the GS energy $\mathcal{E}_0$ increasing monotonically with SP strength, while the excited energy levels (blue lines) fan out progressively as $g/U$ increases. This spectral evolution demonstrates the growing energy cost associated with occupying sites away from the potential minimum as the SP steepens. The many-body spectrum exhibits characteristic features of Wannier-Stark localization, with increasing level spacing and band flattening at higher SP strengths.
The lower panel of Fig. \ref{f9} displays the  $\langle r' \rangle$. The calculated $\langle r' \rangle$ curve (blue) exhibits non-monotonic behavior, with an initial sharp rise to approximately 0.37 at small $g/U$, followed by a decline to around 0.25-0.30 for larger SP values. This behavior indicates that the system does not fully transition to either complete quantum chaos or strict integrability across the examined SP range. Instead, the system occupies an intermediate spectral regime, suggesting the presence of partial symmetries or approximate conservation laws that constrain the eigenvalue distribution. The non-monotonic behavior of $\langle r' \rangle$ with a small peak near $g/U \approx 0.2$ and another near $g/U \approx 1.8$ suggests competition between delocalization due to tunneling and localization induced by the potential SP.

\begin{figure}[t]
    \centering
    \includegraphics[width=1\linewidth]{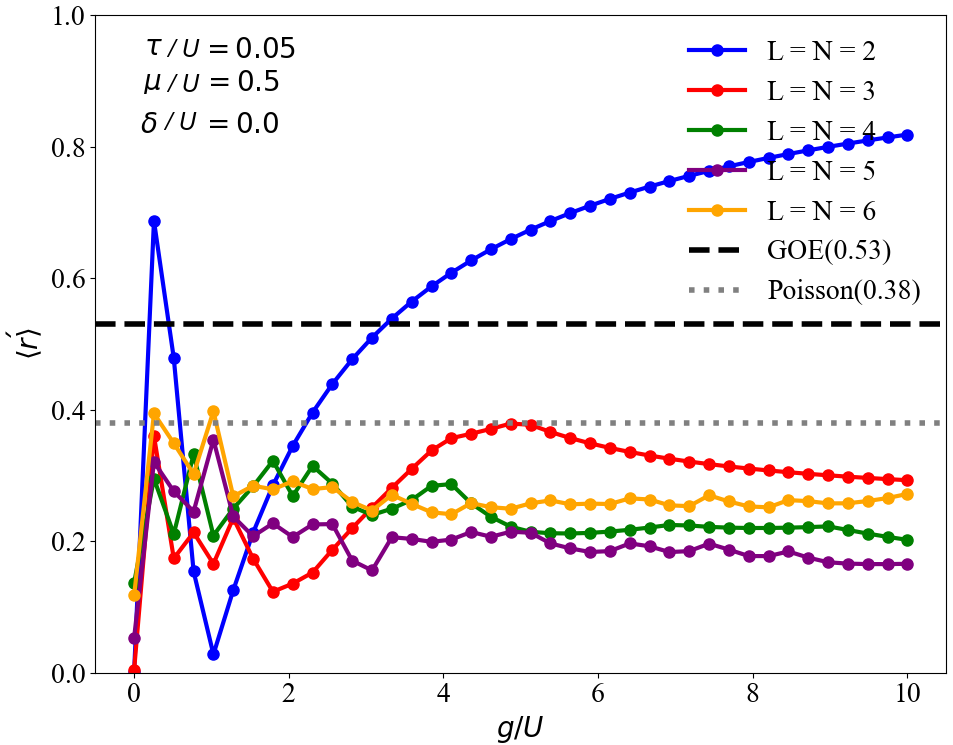}
        \caption{Evolution of $\langle r' \rangle$ versus SP strength $g/U$ for the clean BHM at unit filling ($L = N$), with $\tau /U= 0.05$, $\mu/U = 0.5$, and $\delta/U = 0$. The dashed and dotted lines indicate GOE and Poisson values, respectively. Larger systems remain nonergodic across all $g/U$.
    }
    \label{f10}
\end{figure}

To assess the emergence of ergodicity and quantum chaos in the clean BHM, we compute the $\langle r' \rangle$ as a function of the $g/U$, as shown in Fig.~\ref{f10}.
For the smallest system ($L = N = 2$), we observe that $\langle r' \rangle$ increases with $g/U$ and approaches the GOE value $\langle r' \rangle \approx 0.53$ at large gradients. This trend suggests the emergence of level repulsion and spectral statistics characteristic of quantum chaotic systems. However, this behavior is not sustained at larger system sizes. For $L = 3$, a weak maximum in $\langle r' \rangle$ is seen near $g/U \sim 5$, but its value remains below the GOE limit and subsequently decreases.
As the system size increases ($L = 4, 5, 6$), the values of $\langle r' \rangle$ remain significantly below the GOE threshold and approach or fall below the Poisson value $\langle r' \rangle \approx 0.38$. This trend indicates a lack of level repulsion and is consistent with integrable or localized behavior. Despite the breaking of translational symmetry by the linear gradient, the SP does not induce a sufficient level of mixing to generate chaotic spectral statistics at these sizes and parameters. Instead, the system retains a nonergodic character throughout the range of $g/U$ studied.

\subsection{Role of Disorder}
\begin{figure}[t]
    \centering
    \includegraphics[width=1\linewidth]{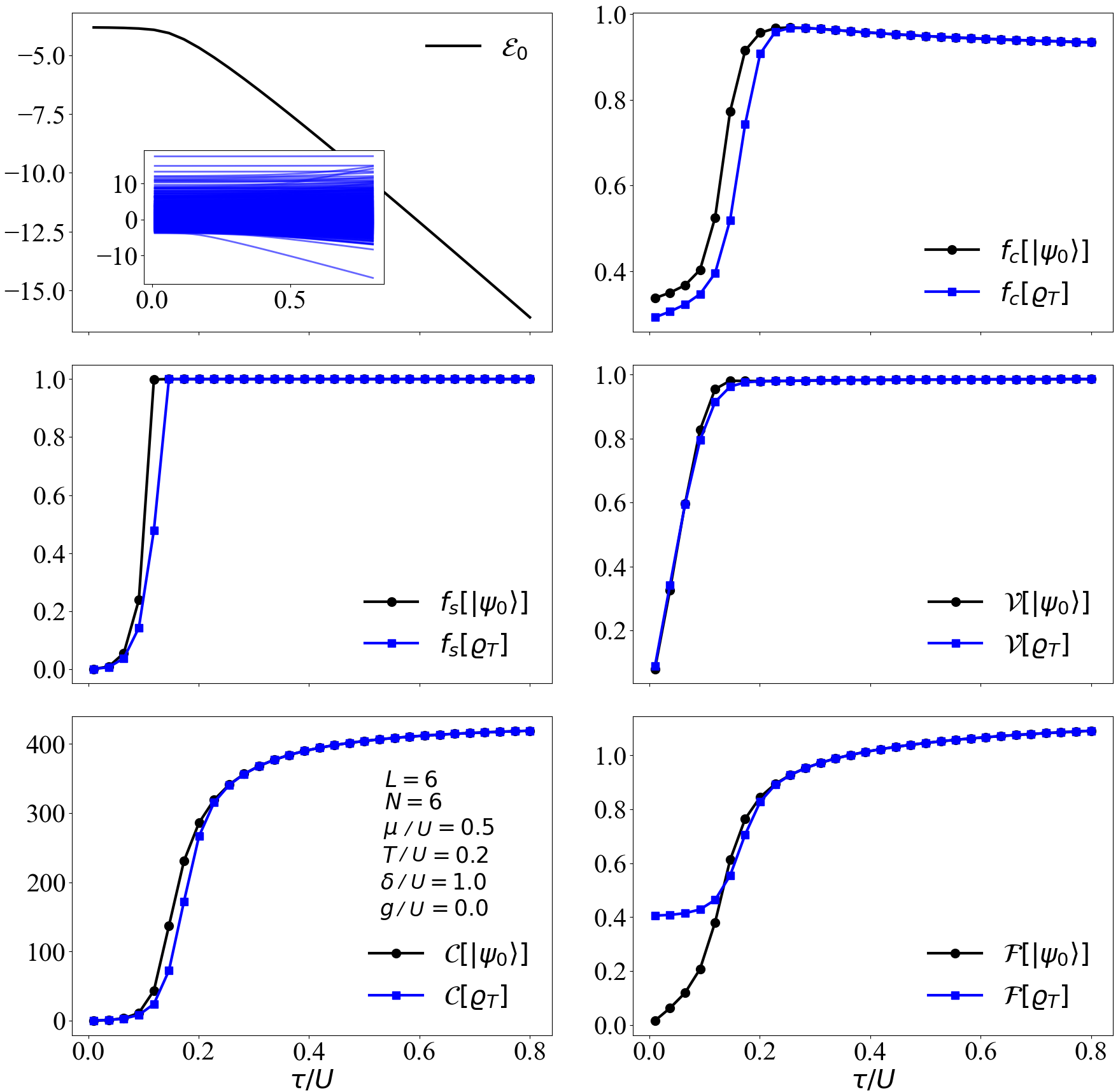}
    \put(-225,220){\textbf{a)}}
    \put(-100,220){\textbf{b)}}
    \put(-227,145){\textbf{c)}}
    \put(-104,145){\textbf{d)}}
    \put(-225,65){\textbf{e)}}
    \put(-102,65){\textbf{f)}}
\caption{
Disordered BHM with fixed values of $\delta/U = 1.0$ and $g/U = 0.0$ at $T/U = 0.2$. GS (black circles) and TS (blue squares) observables versus $\tau/U$ for the unit-filled system. (a) GS energy $\mathcal{E}_0$ with spectrum inset showing irregular spacing indicative of Anderson localization, (b) $f_c$, (c) $f_s$, (d)  $\mathcal{V}$, (e)  $\mathcal{C}$, and (f)  $\mathcal{F}$.
}
\label{figure4}
\end{figure}

\begin{figure}[t]
\centering
\includegraphics[width=1\columnwidth]{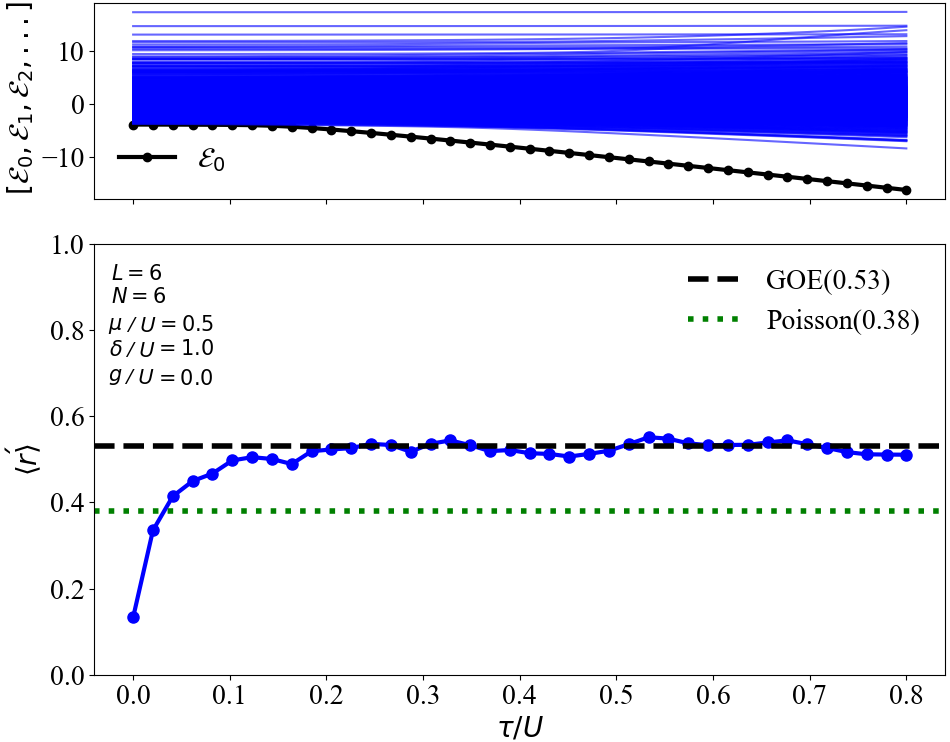}
\caption{(Top) Energy spectrum $\varepsilon_n$ versus $\tau/U$ showing the GS $\varepsilon_0$ (black) and excited states. (Bottom) Level-spacing ratio $\langle r' \rangle$ with GOE (0.53, black dashed) and Poisson (0.38, green dotted) limits. System parameters: $L=6$, $N=6$,  $g/U=0.0$, $\mu/U=0.5$, $\delta/U=1.0$. The persistent separation of $\varepsilon_0$ and suppressed $\langle r' \rangle$ values indicates disorder-stabilized non-ergodic behavior.}
\label{f12}
\end{figure}
We now examine the disordered BHM ($\delta/U = 1.0$, $g/U = 0.0$) at $T/U = 0.2$ for a unit-filled system of $L = N = 6$. Figure~\ref{figure4} compares GS and TS observables versus $\tau/U$.
The GS energy (Fig.~\ref{figure4}(a)) flattens at small $\tau/U$ due to localization-induced mobility suppression. The many-body spectrum reveals highly irregular level spacing without a clear gap, which shows a characteristic of strong Anderson localization, indicating a crossover rather than a sharp phase transition.

The $f_c$ (Fig.~\ref{figure4}(b)) is strongly suppressed at low $\tau/U$ and increases steadily as tunneling dominates over disorder. Thermal effects on $f_c$ are modest except at intermediate $\tau/U$. The $f_s$ (Fig.~\ref{figure4}(c)) shows stronger suppression, remaining near zero until $\tau/U \gtrsim 0.15$, before rising sharply as coherence develops.
The visibility $\mathcal{V}$ shown in Fig.~\ref{figure4}(d) follows a similar trend. Namely it gradually increases upon $\tau/U$ and depicts smooth thermal reduction. The $\ell_1$-norm of coherence $\mathcal{C}$ (Fig.~\ref{figure4}(e)) exhibits thermal enhancement at small $\tau/U$ driven by excited-state contributions, and grows steadily at larger $\tau/U$ values.
Number fluctuations $\mathcal{F}$ (Fig.~\ref{figure4}(f)) mirror $\mathcal{C}$. Namely, they are suppressed in the GS at low $\tau/U$ but enhanced in the TS due to thermal excitations, approaching clean-system values at large $\tau/U$.
Disorder thus suppresses SF and coherence at low $\tau/U$, leading to a smooth crossover rather than a sharp transition. Thermal effects enhance local coherence and fluctuations in the localized regime but become negligible as the tunneling term dominates.

Figure~\ref{f12} shows the evolution of energy levels $\varepsilon_n$ and the corresponding level-spacing ratio $\langle r' \rangle$ as a function of $\tau/U$ for parameters $L=N=6$ with finite disorder $\delta/U=1.0$. Notably, the GS $\varepsilon_0$ (black curve) remains well-separated from excited states throughout the parameter range, suggesting the persistence of localization effects even with increasing hopping strength.

The level statistics show markedly different behavior compared to the clean system case.  $\langle r' \rangle$ begins near the Poisson value $0.38$ in small $\tau/U$, characteristic of localized states in the strong disorder regime, and reaches the GOE limit $0.53$ at some values of $\tau/U$. This incomplete transition toward quantum chaos suggests that the disorder $\delta/U=1.0$ creates an intermediate regime where neither full localization nor complete ergodicity is achieved.

The absence of a SP ($g/U=0.0$) in this configuration allows us to isolate the effects of disorder, which appears to maintain spectral features characteristic of localization despite increasing hopping strength. This contrasts sharply with SP-induced localization, as evidenced by the GS remaining distinctly separated from the excited state continuum. The chemical potential $\mu/U=0.5$ appears to pin the GS energy while leaving the chaotic properties of excited states relatively unaffected.

These results demonstrate how disorder ($\delta/U=1.0$) can suppress the development of quantum chaos even in the presence of significant hopping ($\tau/U \sim 0.8$), suggesting potential pathways for stabilizing non-thermal states in disordered bosonic systems. The intermediate $\langle r' \rangle$ values may indicate a Griffiths-like regime where rare regions dominate the system's behavior.

\begin{figure}[t]
    \centering
    \includegraphics[width=1\columnwidth]{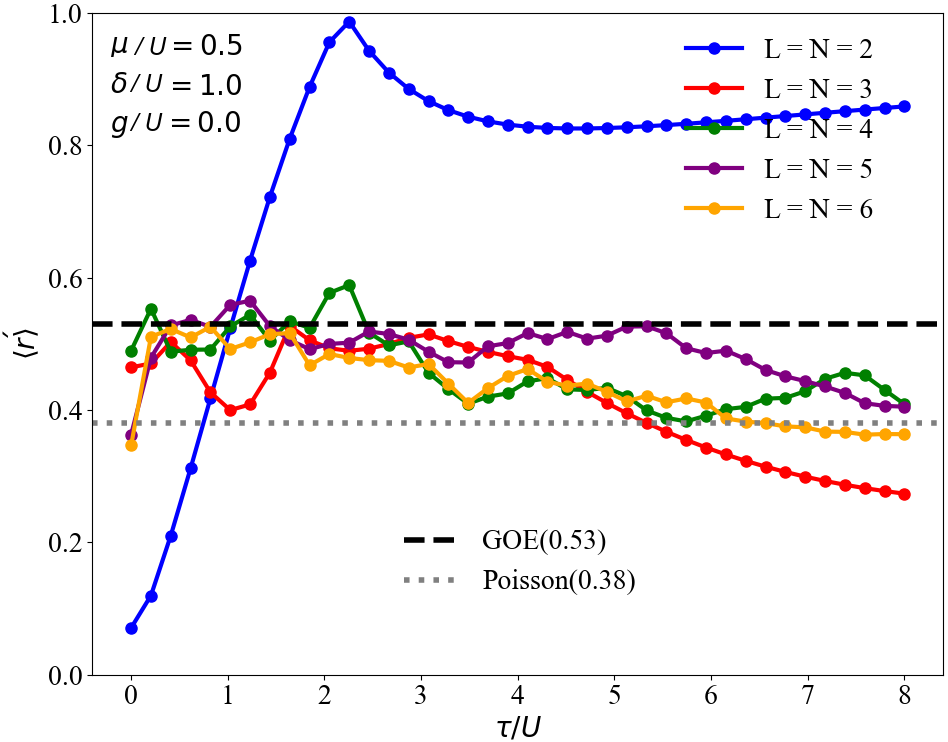}
      \caption{System-size dependence of the MGR $\langle r' \rangle$ versus normalized hopping $\tau/U$ for the BHM with quenched disorder ($g/U = 0.0$, $\mu/U = 0.5$, $\delta/U = 1.0$) at unit filling. Different system sizes ($L = N = 2$ through $L = N = 6$) are shown with their respective colored curves. The dashed and dotted horizontal lines represent the theoretical values for GOE (0.53) and Poisson (0.38) statistics, respectively. The smallest system ($L = N = 2$) shows a dramatic transition from Poisson-like statistics at small $\tau/U$ to a pronounced peak around $\tau/U \approx 2.2$ before settling at elevated values ($\langle r' \rangle \approx 0.85$) at large hopping. Larger systems exhibit more complex behavior, initially oscillating around the GOE value at intermediate hopping strengths before gradually declining toward or below the Poisson value at large $\tau/U$, indicating a return to more integrable dynamics.}
      \label{f13}
\end{figure}

Figure~\ref{f13} illustrates the MGR $\langle r' \rangle$ versus normalized hopping $\tau/U$ for systems with quenched disorder ($\delta/U = 1.0$) across different system sizes while maintaining unit filling. This analysis provides a revealing contrast to our findings for systems with SP (Fig.~\ref{figure4}) and clean systems.

The smallest system ($L = N=2$, blue curve) exhibits striking behavior, with $\langle r' \rangle$ starting near Poisson statistics ($\approx 0.07$) at small $\tau/U$ before sharply rising to a pronounced peak of approximately 0.98 near $\tau/U \approx 2.2$. The MGR then gradually decreases and stabilizes around 0.85 at large hopping strengths. This extreme super-GOE behavior indicates an exceptionally strong level of repulsion in this minimal system when both disorder and significant hopping are present.

Larger systems ($L = N=3,4,5,6$) display more complex spectral evolution. Initially starting near or below Poisson statistics at small $\tau/U$, they quickly rise to oscillate around the GOE value (0.53) for intermediate hopping strengths ($1\lesssim \tau/U \lesssim 3$). However, in stark contrast to the SP case, these larger systems show a gradual decline in $\langle r' \rangle$ as $\tau/U$ increases beyond 3, eventually approaching or falling below the Poisson value (0.38) at large hopping strengths.

The $L = N=3$ system (red curve) exhibits the most dramatic decline, reaching $\langle r' \rangle \approx 0.28$ at $\tau/U = 8$, well below the Poisson limit. Similarly, the $L = N=4$ and $L = N=6$ systems approach the Poisson value at large hopping, while $L = N=5$ maintains slightly higher values but still shows a clear downward trend.
These results reveal that quenched disorder ($\delta/U = 1.0$) fundamentally alters the system's spectral landscape compared to the SP case. While both configurations enable a transition from Poisson-like to GOE-like statistics with initial increases in hopping strength, the quenched disorder appears to drive the system back toward more integrable behavior at very large hopping values for larger system sizes.

This non-monotonic behavior suggests a complex competition between disorder-induced localization and hopping-induced delocalization. At intermediate hopping values, the system exhibits signatures of quantum chaos as the hopping allows particles to overcome the disorder-induced localization. However, at large hopping strengths, the system appears to recover more regular dynamics, possibly due to the dominance of kinetic energy terms that effectively average out the disorder effects, leading to more ballistic transport and reduced level repulsion.

The return to Poisson-like statistics at large $\tau/U$ for extended systems contrasts sharply with the SP case, where larger systems maintained values closer to GOE predictions at high hopping strengths. This difference highlights the distinct impacts of quenched disorder versus SP on the ergodic properties of the BHM and demonstrates the rich complexity of many-body quantum systems subject to different types of spatial inhomogeneities.

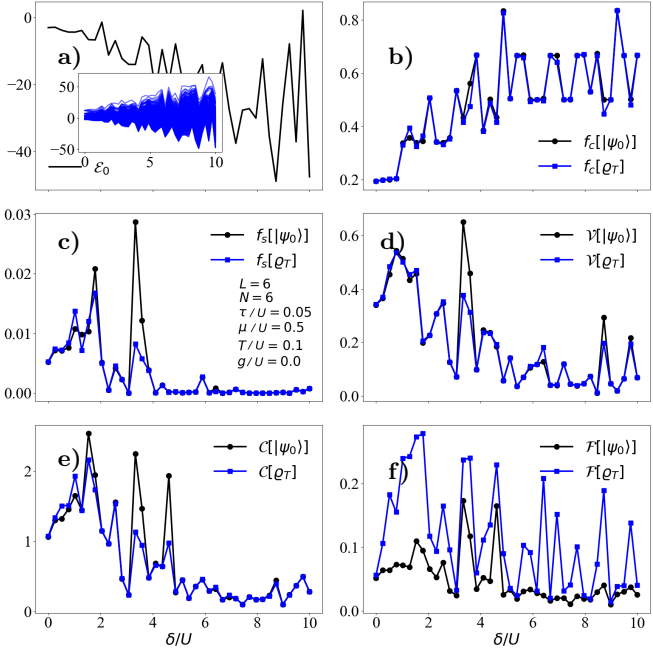
\begin{figure}[t]
    \centering
    \includegraphics[width=1.0\linewidth]{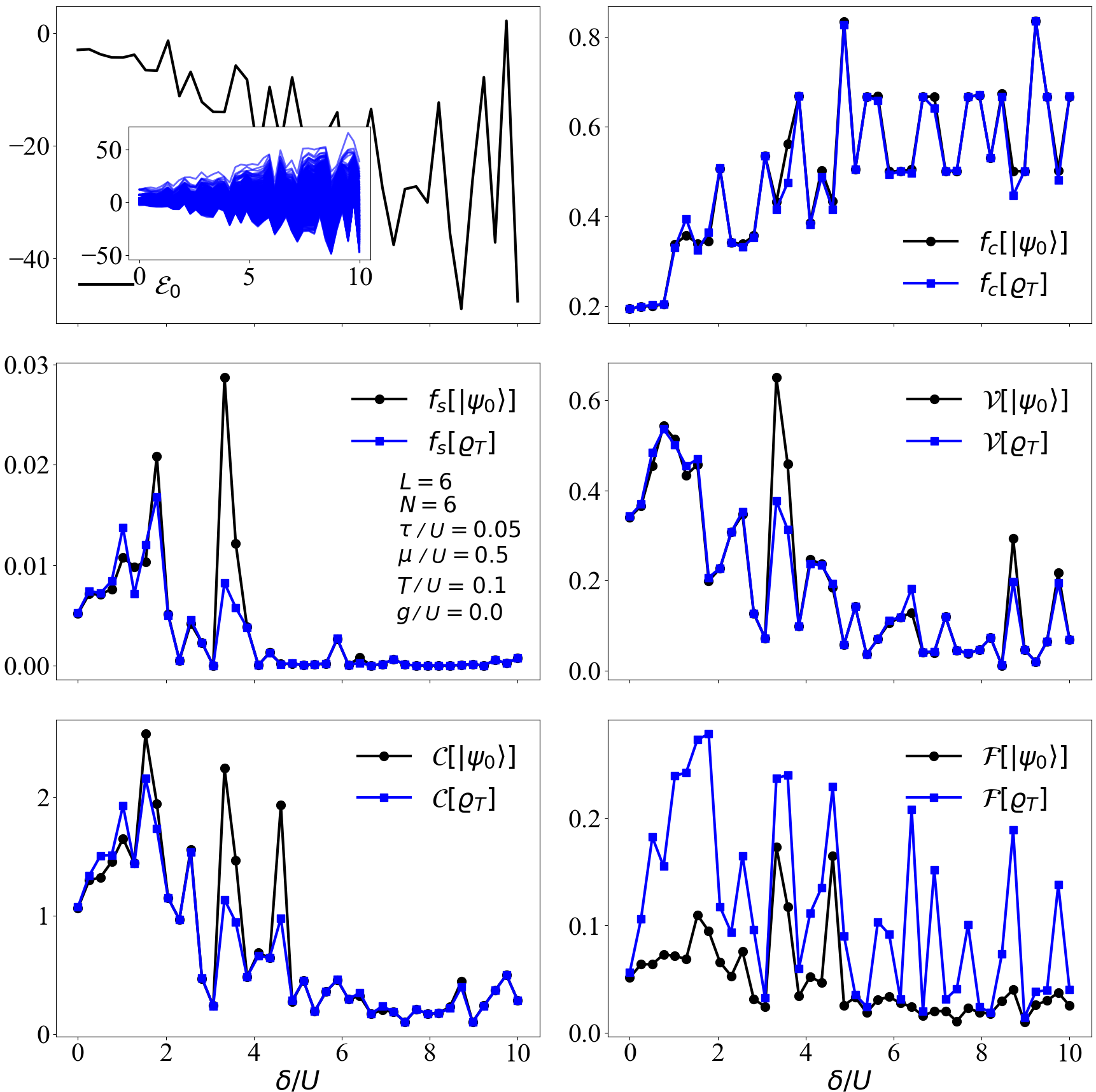}
    \put(-225,220){\textbf{a)}}
    \put(-100,220){\textbf{b)}}
    \put(-225,150){\textbf{c)}}
    \put(-103,150){\textbf{d)}}
    \put(-225,68){\textbf{e)}}
    \put(-100,63){\textbf{f)}}
\caption{Effect of disorder strength $\delta/U$ on the BHM observables with zero SP ($g/U = 0$). GS (black circles) and TS at $T/U = 0.1$ (blue squares). (a) GS energy $\mathcal{E}_0$ with spectrum inset, (b) $f_c$, (c)  $f_s$, (d) 
$\mathcal{V}$, (e)  $\mathcal{C}$, and (f)  $\mathcal{F}$.} %Fixed parameters are: $L = 6$, $N = 6$, $\tau = 0.05$, $U = 1.0$, $\mu = 0.5$.}
    \label{figure5}
\end{figure}

Figure~\ref{figure5} examines how varying disorder strength $\delta$ affects system properties at fixed tunneling $\tau = 0.05$ and interaction strength $U = 1.0$. This parameter choice places the clean system in the MI phase, allowing us to isolate effects due to disorder.

The GS energy (Fig.~\ref{figure5}(a)) increases with disorder strength as the particles encounter energetically unfavorable sites. The many-body spectrum shows progressive level mixing and gap suppression, hallmarks of the transition to an Anderson-localized regime.

All coherence metrics—$f_c$ (Fig.~\ref{figure5}(b)), $f_s$ (Fig.~\ref{figure5}(c)), $\mathcal{V}$ (Fig.~\ref{figure5}(d)), and $\mathcal{C}$ (Fig.~\ref{figure5}(e))—decrease monotonically with increasing $\delta/U$, confirming disorder-induced localization. The SF fraction shows the most dramatic reduction, dropping to near-zero in moderate disorder ($\delta/U \gtrsim 1.0$), while $f_c$ decreases more slowly. This difference highlights that condensation can persist in disconnected regions even when global superfluidity is destroyed. Notably, the TS consistently shows enhanced coherence $\mathcal{C}$ and number fluctuations $\mathcal{F}$ (Fig.~\ref{figure5}(f)) compared to the GS across all disorder strengths. This thermal enhancement becomes proportionally larger at stronger disorder, where thermal excitations activate tunneling between otherwise disconnected localized states, creating short-range coherence absent in the GS.

\begin{figure}[t]
    \centering
    \includegraphics[width=0.95\linewidth]{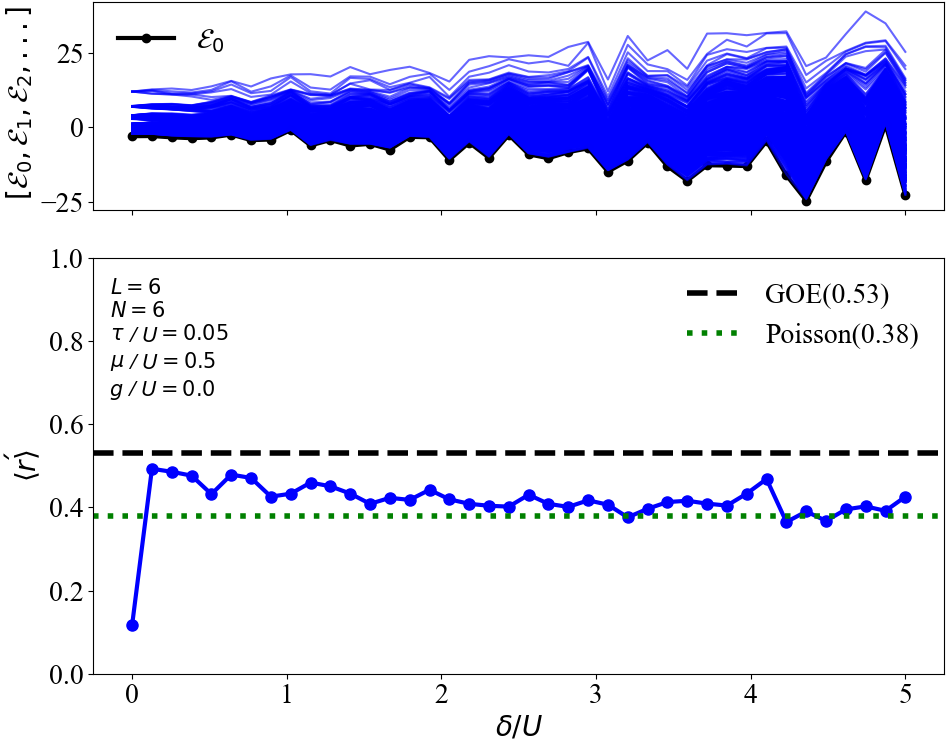}
    \caption{(Top) Many-body energy eigenvalues $\mathcal{E}_n$ of the disordered BHM as a function of disorder strength $\delta/U$, with the GS energy $\mathcal{E}_0$ highlighted in black. (Bottom) Disorder-averaged ratio of adjacent level spacings, $\langle r' \rangle$, compared with the  GOE prediction ($\langle r' \rangle \approx 0.53$, dashed line) and Poisson statistics ($\langle r' \rangle \approx 0.38$, dotted line). Parameters $L = N = 6$, $\tau/U = 0.05$, $g/U = 0$, and $\mu/U = 0.5$ have been assumed.}
    \label{fig:level_statistics}
\end{figure}

To characterize the onset of localization, we analyze the spectral statistics of the disordered BH Hamiltonian by computing the full many-body energy spectrum and evaluating the adjacent level spacing ratio $\langle r' \rangle$, a standard probe of ergodicity breaking. As shown in Fig.~\ref{fig:level_statistics}, the eigenenergy spectrum (top panel) exhibits increased irregularity and bandwidth with increasing disorder strength $\delta/U$, consistent with the emergence of localization. The average ratio $\langle r' \rangle$ (bottom panel) transitions from values near the GOE prediction ($\langle r' \rangle \approx 0.53$), indicative of ergodic level repulsion, toward the Poissonian limit ($\langle r' \rangle \approx 0.38$), characteristic of localized, nonthermal spectra. This crossover, occurring around $\delta/U \approx 1$, signals a breakdown of spectral correlations and provides strong evidence for many-body localization in this interacting system at finite size.

\begin{figure}
    \centering
    \includegraphics[width=1\linewidth]{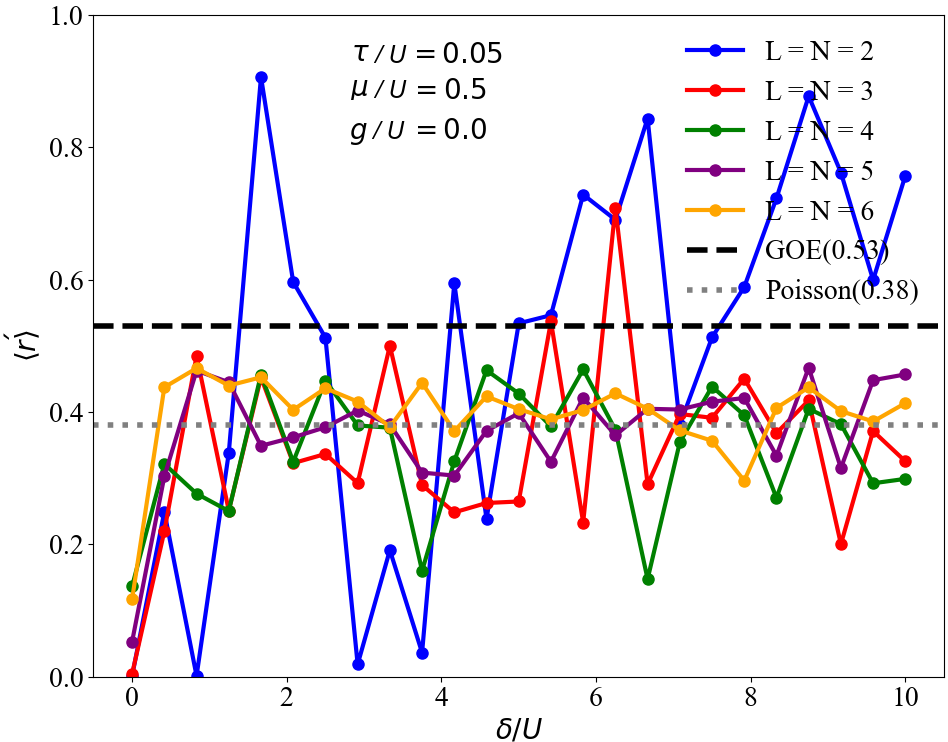}
    \caption{System-size dependence of the MGR $\langle r' \rangle$ versus normalized hopping $\tau/U$ for the BHM with quenched disorder ($g/U = 0.0$, $\mu/U = 0.5$, $\delta/U = 1.0$) at unit filling. Different system sizes $2\leq (L = N) \leq 6$ are shown with their respective colored curves. The dashed and dotted horizontal lines represent the theoretical values for GOE (0.53) and Poisson (0.38) statistics, respectively. The smallest system ($L = N = 2$) shows a dramatic transition from Poisson-like statistics at small $\tau/U$ to a pronounced peak around $\tau/U \approx 2.2$ before settling at elevated values ($\langle r' \rangle \approx 0.85$) at large hopping. Larger systems exhibit more complex behavior, initially oscillating around the GOE value at intermediate hopping strengths before gradually declining toward or below the Poisson value at large $\tau/U$, indicating a return to more integrable dynamics.}
    \label{f16}
\end{figure}

Figure~\ref{f16} illustrates MGR $\langle r' \rangle$ versus normalized hopping $\tau/U$ for systems with quenched disorder ($\delta/U = 1.0$) across different system sizes while maintaining unit filling. This analysis contrasts with our findings for systems with SP (Figure~\ref{figure4}) and clean systems.
The smallest system with $L = N=2$ (blue curve) exhibits striking behavior, with $\langle r' \rangle$ starting near Poisson statistics ($\approx 0.07$) at small $\tau/U$ before sharply rising to a pronounced peak of approximately 0.98 near $\tau/U \approx 2.2$. The MGR then gradually decreases and stabilizes around 0.85 at large hopping strengths. This extreme super-GOE behavior indicates a strong repulsion in this minimal system when disorder and significant hopping are present.

Larger systems ($L = N=3,4,5,6$) in Fig.~\ref{f16} display more complex spectral evolution. Initially starting near or below Poisson statistics at small $\tau/U$, they quickly rise to oscillate around the GOE value (0.53) for intermediate hopping strengths ($1 \lesssim\tau/U \lesssim 3$). However, in stark contrast to the SP case, these larger systems show a gradual decline in $\langle r' \rangle$ as $\tau/U$ increases beyond 3, eventually approaching or falling below the Poisson value (0.38) at large hopping strengths.
The $L = N=3$ system (red curve) exhibits the most dramatic decline, reaching $\langle r' \rangle \approx 0.28$ at $\tau/U = 8$, well below the Poisson limit. Similarly, the $L = N=4$ and $L = N=6$ systems approach the Poisson value at large hopping, while $L = N=5$ maintains slightly higher values but still shows a clear downward trend.
These results reveal that quenched disorder ($\delta/U = 1.0$) fundamentally alters the system's spectral landscape compared to the SP case. While both configurations enable a transition from Poisson-like to GOE-like statistics with initial increases in hopping strength, the quenched disorder appears to drive the system back toward more integrable behavior at very large hopping values for larger system sizes.

This non-monotonic behavior suggests a complex competition between disorder-induced localization and hopping-induced delocalization. At intermediate hopping values, the system exhibits signatures of quantum chaos as the hopping allows particles to overcome the disorder-induced localization. However, at very large hopping strengths, the system appears to recover more regular dynamics, likely due to the dominance of kinetic energy terms that effectively average out the disorder effects, leading to more coherent transport and reduced level repulsion.

The return to Poisson-like statistics at large $\tau/U$ for extended systems contrasts sharply with the SP case, where larger systems maintained values closer to GOE predictions at high hopping strengths. This difference highlights the distinct impacts of quenched disorder versus SP on the ergodic properties of the BHM and demonstrates the rich complexity of many-body quantum systems subject to different types of spatial inhomogeneities.

\section{Conclusions}
\label{sec:conclusion}
% The exact numerical investigation of the 1D BHM uncovers the intricate interplay among thermal fluctuations, SP, and quenched disorder in shaping quantum coherence and phase behavior. We employed multiple complementary observables and studied the complex landscape of quantum phases and transitions under their mutual influence. In the disorder-free system, we have observed an indication of the quantum phase transition at $\tau/U \approx 0.17$, where all coherence metrics, including condensate fraction, SF fraction, visibility, and the $\ell_1$ norm of coherence, increase sharply, marking the transition from a MI to an SF. Finite temperature transforms this abrupt transition into a smooth crossover. Notably, the $\ell_1$-norm and number fluctuations are enhanced in the intermediate regime due to thermal occupation of delocalized excited states, revealing that thermal effects can activate coherence pathways absent in the GS.
Our exact numerical study of the 1D BHM reveals the intricate interplay between thermal fluctuations, a SP, and quenched disorder in shaping quantum coherence and phase behavior. By employing a suite of complementary observables, we examined the complex landscape of quantum phases and transitions arising from their combined effects.
In the absence of disorder, our results indicate a QPT near $\tau/U \approx 0.17$, where all coherence indicators including condensate fraction, SF fraction, visibility, and the $\ell_1$-norm of coherence, exhibit a sharp increase, indicating the transition from a MI to a SF phase. At finite temperatures, this transition is transformed to a smooth crossover. Interestingly, both the $\ell_1$-norm and number fluctuations are significantly enhanced in the intermediate regime, driven by thermal activation of delocalized excited states. This demonstrates that thermal effects can enable coherence mechanisms that are absent in the GS.

Introducing a linear tilt fundamentally alters phase behavior through Wannier-Stark localization. This leads to a delayed onset of superfluidity 
%($\tau/U \gtrsim 0.3$)
lower $f_c$ saturation values, and non-monotonic coherence variations with field strength. We observe resonance-enhanced coherence at specific values of the SP, where inter-site energy offsets match tunneling amplitudes, allowing coherent tunneling despite strong localization. These resonances highlight how external parameters can be tuned to engineer and manipulate coherence.

In disordered systems, Anderson localization dominates, where superfluidity and condensation are suppressed, turning the sharp transition into a gradual crossover. While the SF fraction vanishes at moderate disorder ($\delta /U\gtrsim 1.0$), the condensate fraction decreases gradually. This indicates the persistence of localized condensate "puddles" lacking global phase coherence—signatures of the BG phase. Across all disorder strengths, thermal states exhibit consistently enhanced local coherence and number fluctuations compared to GSs, as thermal excitations enable tunneling between otherwise disconnected regions.

% Our multi-observable analysis shows that while both SP and disorder suppress global coherence, they do so via distinct mechanisms, such as energy gradients versus random scattering, resulting in unique signatures in correlation observables. The $\ell_1$-norm of coherence proves sensitive in detecting subtle coherence structures invisible to traditional observables. Particularly, in the BG regime, where thermal delocalization enhances coherence in excited states.
Our multi-observable analysis reveals that while both the SP and disorder suppress global coherence, they do so through fundamentally distinct mechanisms, namely, energy gradients in the case of the SP and random scattering due to disorder. These distinct origins lead to unique signatures in correlation-based observables. Notably, the $\ell_1$-norm of coherence emerges as a particularly sensitive probe, capable of detecting subtle coherence structures hidden from conventional measures. This sensitivity is especially evident in the BG regime, where thermal delocalization enhances coherence among excited states.

These findings reveal the multiplex interplay among tunneling, interactions, disorder, and temperature in phase transition and quantum properties of the strongly correlated bosonic systems. They hold significant implications for quantum simulation platforms based on ultracold atoms. By elucidating how various perturbations reshape the coherence landscape, our study contributes to a deeper understanding of strongly correlated phases beyond the GS. Furthermore, the combined effects of SP and disorder introduce a richer spectrum of physical phenomena, particularly through the coexistence of deterministic and stochastic localization mechanisms in 1D BH systems.

\vspace{0.5cm}

\noindent \textbf{Acknowledgments:} H.A.Z. acknowledges the financial support provided under
the postdoctoral fellowship program of P. J. Šafárik University in Košice, Slovakia. H.A.Z. is also supported by the grants of the Slovak Research and Development Agency under contract No. APVV-20-0150 and the Ministry of Education, Research,
Development and Youth of the Slovak Republic under contract No. VEGA 1/0298/25.  

\noindent \textbf{Data availability:} All data generated or analyzed during this study are included in this published article.\\
\\
\\
\noindent \textbf{Competing interests:} The authors declare no competing interests.

\appendix
\bibliography{bibliography} 
\end{document}